\documentclass[iop]{emulateapj}

\usepackage{apjfonts} % Times fonts

\usepackage{graphicx}
\usepackage{verbatim}
\usepackage{parskip}
\usepackage{datetime}
\newcommand{\Fermi}{{\it{}Fermi}\ }
\newcommand{\n}{\nodata}
\newcommand{\be}{\begin{itemize}}
\newcommand{\ee}{\end{itemize}}
\newcommand{\muasyr}{\hbox{$\; \mu{\rm as \ y}^{-1}\;$}}
\newcommand{\masyr}{\hbox{$\; {\rm mas \ y}^{-1}\;$}}

\newlength\mystoreparindent

\def\fermi{\textit{Fermi }}

\makeindex
\citeindextrue

\received{January 20, 2019}
\accepted{February 18, 2019}
\submitted{}
\journalinfo{Astrophysical~Journal}
\shorttitle{MOJAVE. XVII. Kinematics \& Parent Population}
\shortauthors{M. L. Lister et al.}

\begin{document}

\title{MOJAVE. XVII. Jet Kinematics and Parent Population Properties
  of Relativistically Beamed Radio-Loud Blazars}

\author{ M. L. Lister\altaffilmark{1},
D. C. Homan\altaffilmark{2},
T. Hovatta\altaffilmark{3,4},
K. I. Kellermann\altaffilmark{5},
S. Kiehlmann\altaffilmark{6},
Y. Y. Kovalev\altaffilmark{7,8,9},
W. Max-Moerbeck\altaffilmark{10},
A. B. Pushkarev\altaffilmark{11,7},
A. C. S. Readhead\altaffilmark{6},
E. Ros\altaffilmark{9},
T. Savolainen\altaffilmark{12,4,9} 
}

\altaffiltext{1}{
Department of Physics and Astronomy, Purdue University, 525 Northwestern Avenue,
West Lafayette, IN 47907, USA;
%\email{mlister@purdue.edu}
}
\altaffiltext{2}{
Department of Physics, Denison University, Granville, OH 43023, USA;
}

\altaffiltext{3}{
Finnish Centre for Astronomy with ESO (FINCA), University of Turku,
FI-20014, Turku, Finland;
}

\altaffiltext{4}{
Aalto University Mets\"ahovi Radio Observatory, Mets\"ahovintie 114,
FI-02540 Kylm\"al\"a, Finland;
}

\altaffiltext{5}{
National Radio Astronomy Observatory, 520 Edgemont Road,
Charlottesville, VA 22903, USA;
}

\altaffiltext{6}{
Owens Valley Radio Observatory, California Institute of Technology,
Pasadena, CA 91125, USA;
}

\altaffiltext{7}{
Astro Space Center of Lebedev Physical Institute,
Profsoyuznaya 84/32, 117997 Moscow, Russia;
}

\altaffiltext{8}{
Moscow Institute of Physics and Technology,
  Dolgoprudny, Institutsky per. 9, Moscow region, 141700, Russia;
}

\altaffiltext{9}{
Max-Planck-Institut f\"ur Radioastronomie, Auf dem H\"ugel 69,
53121 Bonn, Germany;
}
\altaffiltext{10}{
Departamento de Astronomía, Universidad de Chile, Camino El
Observatorio 1515, Las Condes, Santiago, Chile;
}

\altaffiltext{11}{
Crimean Astrophysical Observatory, 98409 Nauchny, Crimea, Russia;
}

\altaffiltext{12}{ Aalto University Department of Electronics and
  Nanoengineering, PL 15500, FI-00076 Aalto, Finland;
}

\begin{abstract}
  We present results from a parsec-scale jet kinematics study of 409
  bright radio-loud AGNs based on 15 GHz VLBA data obtained between
  1994 August 31 and 2016 December 26 as part of the 2cm VLBA survey
  and MOJAVE programs. We tracked 1744 individual bright features in
  382 jets over at least five epochs.  A majority (59\%) of the
  best-sampled jet features showed evidence of accelerated motion at
  the $>3 \sigma$ level. Although most features within a jet typically
  have speeds within $\sim 40\%$ of a characteristic median value, we
  identified 55 features in 42 jets that had unusually slow pattern
  speeds, nearly all of which lie within 4 pc (100 pc de-projected) of
  the core feature. Our results combined with other speeds from the
  literature indicate a strong correlation between apparent jet speed
  and synchrotron peak frequency, with the highest jet speeds being
  found only in low-peaked AGNs.  Using Monte Carlo simulations, we
  find best fit parent population parameters for a complete sample of
  174 quasars above 1.5 Jy at 15 GHz. Acceptable fits are found with a
  jet population that has a simple unbeamed power law luminosity
  function incorporating pure luminosity evolution, and a power law
  Lorentz factor distribution ranging from 1.25 to 50 with slope $-1.4
  \pm 0.2$.  The parent jets of the brightest radio quasars have a
  space density of $261 \pm 19$ Gpc$^{-3}$ and unbeamed 15 GHz
  luminosities above $\sim 10^{24.5}$ W Hz$^{-1}$, consistent with FR
  II class radio galaxies.

\end{abstract}

\section{INTRODUCTION} 

Relativistic jets from active galactic nuclei (AGN) represent some of
the most energetic known phenomena in the universe, and played a key
role in regulating galaxy formation at early epochs via feedback
processes \citep{2018arXiv181206025B}. One of the most powerful tools
for investigating these outflows is the Very Long Baseline Array
(VLBA), which can be used to provide full polarization,
sub-milliarcsecond scale imaging at radio wavelengths.

Since the VLBA's inauguration in 1994, we have carried out a long term
program to investigate the parsec-scale properties of several hundred
of the brightest AGN jets in the northern sky. This effort started out
as the 2cm VLBA survey \citep{1998AJ....115.1295K}, and continued as
the MOJAVE survey in 2002 with the addition of full polarization
imaging of a complete flux density-limited sample. We have presented
the results from MOJAVE in a number of papers in this series,
including our most recent  analysis of jet kinematics based on
multi-epoch data obtained between 1994 August 31 and 2013 August 20
\citep{MOJAVE_XIII}. 

In this paper we perform a new kinematics analysis that adds
VLBA data taken up to 2016 December 26, and extends the number of AGN
jets studied from 274 to 409.  Most of the new AGNs were added to the
MOJAVE program based on their detection in GeV gamma-rays by the LAT
instrument aboard the \fermi observatory. We also update and expand
our 1.5 Jy flux density-limited sample from 181 to 230 AGNs based on
data from the RATAN 600m telescope and OVRO 40m telescope monitoring
observations at 15 GHz. This sample is now the
largest and most complete radio-loud blazar sample to date, covering
75\% of the entire sky. Using Monte Carlo simulations, we deconvolve
the effects of Doppler boosting and Malmquist bias in this sample to
uncover the intrinsic jet properties of the bright radio loud quasar
population.

The layout of the paper is as follows. In Section~\ref{data} we
describe our VLBA observations and the new flux density-limited 1.5 Jy
Quarter Century (1.5JyQC) AGN sample. We describe our Gaussian fitting
of bright jet features and their apparent trajectories, and discuss
our general findings on the parsec-scale jet kinematics of our sample
in Section~\ref{analysis}. In Section~\ref{montecarlo} we describe the
best fit parent population properties for 174 quasars in the 1.5JyQC
sample based on Monte Carlo simulations.  We use this best fit
simulation to describe the likely viewing angle, Lorentz factor, and
Doppler factor distributions of bright radio-loud quasars.

Throughout this paper we adopt the convention $S_\nu \propto
\nu^{\alpha}$ for spectral index $\alpha$, and use the cosmological
parameters $\Omega_m = 0.27$, $\Omega_\Lambda = 0.73$ and $H_o = 71 \;
\mathrm{km\; s^{-1} \; Mpc^{-1}}$ \citep{Komatsu09}.

\section{OBSERVATIONAL DATA}
\label{data}

The observational data set consists of 15 GHz VLBA observations of 409
AGNs obtained between 1994 August 31 and 2016 December 26 as part of
the MOJAVE program, with supplementary data from the NRAO archive.
These AGNs all have 15 GHz flux density $\gtrsim 0.1$ Jy, and have at
least 5 VLBA epochs spaced in time.  The epoch coverage and cadence
varies considerably among the AGNs as they are members or candidate
members of various radio and gamma-ray selected samples that have been
added at various stages of the program (see \citealt{MOJAVE_XV}).  We
have previously presented the VLBA total intensity and polarization
images in \cite{MOJAVE_I}, \cite{MOJAVE_V}, \cite{MOJAVE_X},
\cite{MOJAVE_XIII}, and \cite{MOJAVE_XV}. These images are also
available from our online data
archive\footnote{\url{http://www.astro.purdue.edu/MOJAVE}}.  We obtained
observer frame values for the low energy (synchrotron) peak frequency
from the literature or via the ASDC spectral energy distribution (SED)
builder \citep{2011arXiv1103.0749S}. We list the overall properties of
the AGNs in Table~\ref{gentable} and Table~\ref{gentable1.5}. The
latter contains 19 AGNs that have not been observed in the MOJAVE VLBA
program, but are new additions to the new 1.5 Jy sample, as we
describe in the next section.
\subsection{\label{MOJAVE_1_5QC}The MOJAVE 1.5 Jy Quarter Century Sample}

\begin{deluxetable*}{lllccccll} 
 \tablecolumns{9} 
% \rotate 
% \tabletypesize{\scriptsize} 
 \tablewidth{0pt}  
 \tablecaption{\label{gentable} AGN Properties}  
 \tablehead{ \colhead{B1950}  & \colhead {Alias} & \colhead{Opt.} & \colhead{$z$}& \colhead{log  $\nu_\mathrm{p}$} &\colhead{Ref.}
  &\colhead{$\mu_\mathrm{max}$} &\colhead{$\beta_\mathrm{max}$}  &\colhead{Reference}  \\ 
 \colhead{(1)} & \colhead{(2)} & \colhead{(3)} & \colhead{(4)} & \colhead{(5)} & 
  \colhead{(6)} & \colhead{(7)} & \colhead{(8)} & \colhead{(9)} } 
 \startdata 

 0003+380\tablenotemark{a} &  S4 0003+38 & Q& 0.229 & 13.1 &10 & 317 $\pm$ 25  & 4.61 $\pm$ 0.36 &  \cite{1994AAS..103..349S} \\ 
 0006+061\tablenotemark{a} &  TXS 0006+061 & B& \n & 13.4 &10 & 221 $\pm$ 43  & \n &  \n \\ 
 0011+189\tablenotemark{a} &  RGB J0013+191 & B& 0.477 & 13.7 &1 & 159 $\pm$ 16  & 4.54 $\pm$ 0.46 &  \cite{2013ApJ...764..135S} \\ 
 0010+405 &  4C +40.01 & Q& 0.256 & 12.9 &1 & 428 $\pm$ 40  & 6.92 $\pm$ 0.64 &  \cite{1992ApJS...81....1T} \\ 
 0015$-$054\tablenotemark{a} &  PMN J0017$-$0512 & Q& 0.226 & 13.6 &10 & 50 $\pm$ 20  & 0.72 $\pm$ 0.28 &  \cite{2012ApJ...748...49S} \\ 
 0019+058\tablenotemark{a} &  PKS 0019+058 & B& \n & 13.1 &10 & 257 $\pm$ 35  & \n &  \cite{2013ApJ...764..135S} \\ 
 0027+056 &  PKS 0027+056 & Q& 1.317 & 12.4 &1 & 22.7 $\pm$ 5.9  & 1.45 $\pm$ 0.38 &  \cite{1999AJ....117...40S} \\ 
 0026+346 &  B2 0026+34 & G& 0.517 & \n &\n & 57 $\pm$ 23  & 1.76 $\pm$ 0.70 &  \cite{2002AJ....124..662Z} \\ 
 0035+413 &  B3 0035+413 & Q& 1.353 & 12.3 &1 & 113.8 $\pm$ 4.7  & 7.40 $\pm$ 0.31 &  \cite{1993AAS..101..521S} \\ 
 0044+566\tablenotemark{a} &  GB6 J0047+5657 & B& 0.747 & \n &\n & 24.7 $\pm$ 6.7  & 1.03 $\pm$ 0.28 &  \cite{2005ApJ...626...95S} \\ 
 0048$-$071\tablenotemark{a} &  OB $-$082 & Q& 1.975 & 12.8 &10 & 131 $\pm$ 10  & 10.79 $\pm$ 0.85 &  \cite{1983MNRAS.205..793W}
\enddata 
\tablecomments{Columns are as follows: 
(1) B1950 name, 
(2) other name,
(3) optical classification, where B = BL Lac, Q = quasar, G = radio galaxy, N = narrow-line Seyfert 1, and U = unknown spectral class, 
(4) redshift,
(5) log of observer frame synchrotron peak frequency in Hz,
(6) reference for synchrotron peak frequency measurement,
(7) maximum jet speed in $\mu$as y$^{-1}$,
(8) maximum jet speed in units of the speed of light,
(9) reference for redshift and/or optical classification.
Reference codes for synchrotron peak frequency measurements:
1. {ASDC SED builder}
2. {\cite{2011ApJ...740...98M}}
3. {\cite{2008AA...488..867N} }
4. {\cite{2011ApJ...743..171A}}
5. {\cite{2006AA...445..441N} }
6. {\cite{2009ApJ...707...55A}}
7. {\cite{2009ApJ...707L.142A}}
8. {\cite{2015AA...578A..69H} }
9. {\cite{2015AA...578A..69H} }
10. {\cite{2015ApJ...810...14A}}
11. {\cite{2015MNRAS.450.3568X}}
12. {\cite{2017AA...598A..17C} }
13. {\cite{2017ApJS..232...18A}}
}

\tablenotetext{a}{Known association with Fermi-LAT gamma-ray source.}
\tablenotetext{b}{Known TeV gamma-ray emitter (http://tevcat.uchicago.edu).}
\tablenotetext{c}{Speed measurement from \cite{2010ApJ...723.1150P}.}

\tablenotetext{}{(This table is available in its entirety in machine-readable form.)}

\end{deluxetable*} 
 
\begin{deluxetable*}{lllcccccll} 
 \tablecolumns{10} 
% \rotate 
% \tabletypesize{\scriptsize} 
 \tablewidth{0pt}  
 \tablecaption{\label{gentable1.5}MOJAVE 1.5 Jy Quarter Century AGN Sample Properties}
 \tablehead{ \colhead{B1950}  & \colhead {Alias} & \colhead{Opt.} & \colhead{$z$} 
  &\colhead{$S_\mathrm{max}$}& \colhead{log  $\nu_\mathrm{p}$} &\colhead{Ref.}
  &\colhead{$\mu_\mathrm{max}$} &\colhead{$\beta_\mathrm{max}$}  &\colhead{Reference}  \\ 
 \colhead{(1)} & \colhead{(2)} & \colhead{(3)} & \colhead{(4)} & \colhead{(5)} & 
  \colhead{(6)} & \colhead{(7)} & \colhead{(8)} & \colhead{(9)} & \colhead{(10)}  } 
 \startdata 

 0003$-$066 &  NRAO 005 & B& 0.347 & 5.33 & 13.0 &1 & 330.4 $\pm$ 9.7  & 7.08 $\pm$ 0.21 &  \cite{2005PASA...22..277J} \\ 
 0007+106 &  III Zw 2 & G& 0.089 & 2.25 & 13.3 &1 & 269 $\pm$ 50  & 1.58 $\pm$ 0.29 &  \cite{1970ApJ...160..405S} \\ 
 0016+731 &  S5 0016+73 & Q& 1.781 & 3.78 & 12.3 &1 & 98.5 $\pm$ 4.1  & 7.64 $\pm$ 0.32 &  \cite{1986AJ.....91..494L} \\ 
 0048$-$097\tablenotemark{a} &  PKS 0048$-$09 & B& 0.635 & 2.33 & 14.3 &1 & \n  & \n &  \cite{2012AA...543A.116L} \\ 
 0059+581\tablenotemark{a} &  TXS 0059+581 & Q& 0.644 & 5.98 & 12.7 &1 & 233.2 $\pm$ 9.1  & 8.62 $\pm$ 0.34 &  \cite{2005ApJ...626...95S} \\ 
 0106+013\tablenotemark{a} &  4C +01.02 & Q& 2.110 & 4.31 & 12.5 &1 & 300 $\pm$ 21  & 25.6 $\pm$ 1.8 &  \cite{LAMOSTDR4} \\ 
 0109+224\tablenotemark{a,b} &  S2 0109+22 & B& \n & 1.50 & 13.4 &1 & 10.7 $\pm$ 4.0  & \n &  \cite{2017ApJ...837..144P} \\ 
 0109+351 &  B2 0109+35 & Q& 0.450 & 1.53 & 12.8 &1 & 198 $\pm$ 54  & 5.4 $\pm$ 1.5 &  \cite{1996MNRAS.282.1274H} \\ 
 0113$-$118\tablenotemark{a} &  PKS 0113$-$118 & Q& 0.671 & 1.87 & 12.9 &10 & 449 $\pm$ 45  & 17.2 $\pm$ 1.7 &  \cite{2012ApJ...748...49S} \\ 
 0119+115 &  PKS 0119+11 & Q& 0.571 & 4.36 & 12.7 &1 & 557 $\pm$ 25  & 18.61 $\pm$ 0.82 &  \cite{2017AA...597A..79P} \\ 
 0122$-$003 &  UM 321 & Q& 1.076 & 1.62 & 12.7 &1 & 252 $\pm$ 58  & 14.0 $\pm$ 3.2 &  \cite{2010AJ....139.2360S}
\enddata 
\tablecomments{Columns are as follows: 
(1) B1950 name, 
(2) other name,
(3) optical classification, where B = BL Lac, Q = quasar, G = radio galaxy, N = narrow-line Seyfert 1, and U = unknown spectral class, 
(4) redshift,
(5) maximum 15 GHz VLBA flux density in Jy between 1994.0 and 2019.0,
(6) log of observer frame synchrotron peak frequency in Hz,
(7) reference for synchrotron peak frequency measurement,
(8) maximum jet speed in $\mu$as y$^{-1}$,
(9) maximum jet speed in units of the speed of light,
(10) reference for redshift and/or optical classification.
Reference codes for synchrotron peak frequency measurements:
1. {ASDC SED builder}
2. {\cite{2011ApJ...740...98M}}
3. {\cite{2008AA...488..867N} }
4. {\cite{2011ApJ...743..171A}}
5. {\cite{2006AA...445..441N} }
6. {\cite{2009ApJ...707...55A}}
7. {\cite{2009ApJ...707L.142A}}
8. {\cite{2015AA...578A..69H} }
9. {\cite{2015AA...578A..69H} }
10. {\cite{2015ApJ...810...14A}}
11. {\cite{2015MNRAS.450.3568X}}
12. {\cite{2017AA...598A..17C} }
13. {\cite{2017ApJS..232...18A}}
}

\tablenotetext{a}{Known association with Fermi-LAT gamma-ray source.}
\tablenotetext{b}{Known TeV gamma-ray emitter (http://tevcat.uchicago.edu).}
\tablenotetext{c}{Speed measurement from \cite{2017ApJ...846...98J}.}

\tablenotetext{}{(This table is available in its entirety in machine-readable form.)}

\end{deluxetable*}

In \cite{2011ApJ...742...27L} and \cite{MOJAVE_X}, we compiled the
MOJAVE 1.5 Jy sample, which consists of all AGNs north of J2000
declination $-30^\circ$ known to have exceeded 1.5 Jy in 15 GHz VLBA
correlated flux density between 1994.0 and 2010.0.  We used a 16 year
selection period in order to include low-duty cycle AGNs that may only
exceed 1.5 Jy for short durations. Despite this, the number counts of
the sample as a function of flux density suggested some incompleteness
below $\sim 1.8$ Jy.  For this reason, we have now extended the
selection period to encompass 25 years (1994.0--2019.0), and use the
extensive 15 GHz OVRO \citep{2011ApJS..194...29R}, RATAN 600m
\citep{2002PASA...19...83K} and 14.5 GHz UMRAO
\citep{1985ApJS...59..513A} monitoring databases to identify
additional AGNs meeting our selection criteria. We estimated the VLBA
flux density from these single-dish measurements by establishing the
amount of extended arcsecond-scale emission with near-simultaneous
VLBA measurements of each AGN at at least one epoch.  This emission is
invisible to the VLBA and is typically non-variable due to its large
size scale. In the case of a small number of AGNs where no
simultaneous measurements were available, we checked the VLA
calibrator list, radio spectra, and published VLA images to verify
that they had no significant arcsecond-scale emission. During this
process, we obtained a better arcsecond-scale emission measurement for
the original 1.5 Jy sample member TXS 0730$+$504, and found that its
maximum inferred VLBA flux density no longer exceeded 1.5 Jy.

The new MOJAVE 1.5 Jy quarter century sample (1.5JyQC;
Table~\ref{gentable1.5}) contains 177 quasars, 38 BL Lac objects, 10
radio galaxies, 1 narrow-line Seyfert 1 galaxy, and 6 AGNs with no
optical spectroscopic information. Of these 232 AGNs, 19 have not
been observed in the MOJAVE or 2cm VLBA survey programs. The redshift
information on the sample is 91\% complete, and 177 (76\%) of the AGNs
have been reported in the literature as associations for gamma-ray
sources detected by the LAT instrument on board the \fermi satellite.

\section{DATA ANALYSIS}
\label{analysis}
\subsection{Gaussian Modeling}

The median redshift of the 409 AGNs analyzed in this paper is $z
\simeq 0.9$, which translates into a spatial scale of $\sim 8$ pc
mas$^{-1}$. The VLBA has an angular resolution at 15 GHz of 0.5 mas to
1 mas (depending on image weighting and target declination), and in
our snapshot mode observations (several scans at different hour
angles, with a total integration time of 30--50 minutes), emission can
usually be detected only out to a few milliarcseconds from the base of
the jet. Any fine-scale sub-pc structure can  be probed only
in the nearest ($z \lesssim 0.1$) AGNs, which comprise fewer than 7\%
of our sample. For this reason, the emission structure of most of the
jets can be well-modeled by a small number of features having a
two-dimensional Gaussian or delta-function intensity profile.

We modeled the sky brightness distribution for each VLBA observation in the
$(u,v)$ visibility plane using the {\it modelfit} task in the Difmap
software package \citep{DIFMAP}. We list the properties of the fitted
features in Table~\ref{gaussiantablestub}.  In some instances, it was
impossible to robustly cross-identify the same features in a jet
from one epoch to the next. We indicate the features with robust
cross-identifications across at least five epochs in column 10 of
Table~\ref{gaussiantablestub}. For the non-robust features, we caution
that the assignment of the same identification number across epochs
does not necessarily indicate a reliable cross-identification.

\begin{deluxetable*}{lclcrrcrcc} 
\tablecolumns{10} 
%\tabletypesize{\scriptsize} 
\tablewidth{0pt}  
\tablecaption{\label{gaussiantablestub}Fitted Jet Features}  
\tablehead{\colhead{} & \colhead {} &   \colhead {} & 
 \colhead{I} & \colhead{r} &\colhead{P.A.} & \colhead{Maj.} & 
\colhead{} &\colhead{Maj. P.A.}   \\  
\colhead{Source} & \colhead {I.D.} &  \colhead {Epoch} & 
\colhead{(mJy)} & \colhead{(mas)} &\colhead{(\arcdeg)} & \colhead{(mas)} & 
\colhead{Ratio} &\colhead{(\arcdeg)}&\colhead{Robust?}   \\  
\colhead{(1)} & \colhead{(2)} & \colhead{(3)} & \colhead{(4)} &  
\colhead{(5)} & \colhead{(6)} & \colhead{(7)} & \colhead{(8)} & 
 \colhead{(9)} &  \colhead{(10)}} 
\startdata 
0003+380  & 0& 2006 Mar 9  & 489  & 0.04 & 290.7 & 0.23 & 0.33 & 292 & Y\\ 
0003+380  & 1& 2006 Mar 9  & 7.2  & 3.98 & 121.8 & 0.72 & 1 & \n & Y\\ 
0003+380  & 2& 2006 Mar 9  & 42.1  & 1.25 & 110.5 & 0.51 & 1 & \n & Y\\ 
0003+380  & 6& 2006 Mar 9  & 104  & 0.28 & 114.6 & 0.27 & 1 & \n & Y\\ 
0003+380  & 7& 2006 Mar 9  & 2.9  & 2.31 & 119.3 & \n & \n & \n & N\\ 
0003+380  & 0& 2006 Dec 1  & 320  & 0.10 & 308.1 & 0.25 & 0.29 & 295 & Y\\ 
0003+380  & 1& 2006 Dec 1  & 4.8  & 3.65 & 120.8 & 1.63 & 1 & \n & Y\\ 
0003+380  & 2& 2006 Dec 1  & 20.9  & 1.56 & 111.0 & 0.25 & 1 & \n & Y\\ 
0003+380  & 5& 2006 Dec 1  & 22.9  & 0.75 & 116.2 & 0.32 & 1 & \n & Y\\ 
0003+380  & 6& 2006 Dec 1  & 145  & 0.45 & 116.3 & 0.05 & 1 & \n & Y
\enddata 

\tablenotetext{a}{Individual feature epoch not used in kinematic fits.}

\tablecomments{Columns are as follows: (1) B1950 name, (2) feature identification number (zero indicates core feature), (3) observation epoch, (4) flux density at 15 GHz in mJy, (5) position offset from the core feature (or map center for the core feature entries) in milliarcseconds, (6) position angle with respect to the core feature (or map center for the core feature entries) in degrees,  (7) FWHM major axis of fitted Gaussian in milliarcseconds, (8) axial ratio of fitted Gaussian, (9) major axis position angle of fitted Gaussian in degrees, (10) robust feature flag. }

\tablenotetext{}{(This table is available in its entirety in machine-readable form.)}

\end{deluxetable*}

Based on previous analysis \citep{MOJAVE_VI}, we estimate the typical
uncertainties in the feature centroid positions to be $\sim 20$\% of
the FWHM naturally-weighted image restoring beam dimensions. For
isolated bright and compact features, the positional errors are
smaller by approximately a factor of two.  We estimate the formal
errors on the feature sizes to be roughly twice the positional error,
according to \cite{1999ASPC..180..301F}.  The flux density accuracies
are approximately 5\% (see Appendix A of
\citealt{2002ApJ...568...99H}), but can be significantly larger for
features located very close to one another. Also, at some epochs which
lacked data from one or more antennas, the fit errors of some features
are much larger.  We do not use the latter in our kinematics analysis,
and indicate them with flags in Table~\ref{gaussiantablestub}.

\subsection{Jet Kinematics}

As in our previous papers (\citealt{MOJAVE_VI, MOJAVE_VII, MOJAVE_X,
  MOJAVE_XII, MOJAVE_XIII}), we analyze the kinematics of jet features
using three methods: (i) a simple one-dimensional radial motion fit,
(ii) a non-accelerating vector fit in two (sky) dimensions, and (iii)
a constant acceleration fit (for features with ten or more epochs).
We use the radial fit for diagnostic purposes only (see below), and do
not tabulate those fit results here. In all cases, we assume the
bright core feature (id = 0 in Table~\ref{gaussiantablestub}) to be
stationary, and measure the positions of jet features at all epochs
with respect to it.

We have modified our model slightly from our previous papers, and now
fit for the sky position of each feature at a reference middle epoch
$t_\mathrm{mid}$, rather than fitting for the epoch of origin in the
$x$ (right ascension) and $y$ (declination) sky directions. Our new
parametrization is as follows:
\begin{eqnarray}
x(t) &=& x_\mathrm{mid} + \mu_x(t-t_\mathrm{mid}) + \frac{\dot{\mu}_x}{2}(t-t_\mathrm{mid})^2 ,\\
y(t) &=& y_\mathrm{mid} + \mu_y(t-t_\mathrm{mid}) + \frac{\dot{\mu}_y}{2}(t-t_\mathrm{mid})^2, 
\end{eqnarray}

\noindent where $t_\mathrm{mid}$ is the numerical mean of the first
and last observation epoch dates for the feature being fitted, and $\mu_x$
and $\mu_y$ are the fitted angular speeds in each sky direction. For
the vector fits, the accelerations $\dot{\mu}_x$ and $\dot{\mu}_y$ are
fixed to zero, and for the radial motion fits, we used the alternate
parameterization $r(t) = r_\mathrm{mid} +\mu_r(t-t_\mathrm{mid})$,
where $r(t)$ is the radial distance from the core feature at time $t$.

We made radial and vector motion fits using all of the available
data from 1994 August 31 to 2016 December 26 on 1744 robust jet
features in 382 jets. There were 27 jets in which we could not
identify any robust features due to a lack of sufficiently strong
downstream jet flux or a suitably stable core feature, or insufficient
spatial resolution. We are carrying out a followup 43 GHz multi-epoch
VLBA study on several of these jets.

In Table~\ref{vectormotiontablestub} we list the results of the vector
motion fits. Due to the nature of our kinematic model, which naturally
includes the possibility of accelerated motion, we did not estimate
ejection epochs (Column 12) for any features where we could not
confidently extrapolate their motion to the core.  Jet features for
which we list an ejection epoch had the following properties: (i)
significant motion $(\mu \ge 3\sigma_\mu)$, (ii) no significant
acceleration, (iii) a velocity vector direction $\phi$ within $15^\circ$ of the outward radial direction to high confidence, i.e.,
$|\langle\vartheta\rangle - \phi|+2\sigma \le 15^\circ$, where
$\vartheta$ is the mean position angle, (iv) an extrapolated position
at the ejection epoch no more than $0.2$ mas from the core, and (v) a
fitted ejection epoch that differed by no more than 0.5 years from
that given by the radial motion fit.

\begin{deluxetable*}{lcrrrrrrrrrrrrr} 
%\rotate 
\tablecolumns{15} 
%\tabletypesize{\scriptsize} 
\tablewidth{0pt}  
\tablecaption{\label{vectormotiontablestub}Vector Motion Fit Properties of Jet Features}  
\tablehead{\colhead{} & \colhead {} &   \colhead {} & 
\colhead{$\langle S\rangle$}  &\colhead{$\langle R\rangle$} &\colhead{$\langle d_{\mathrm{proj}}\rangle$} & \colhead{$\langle\vartheta\rangle$} & 
 \colhead{$\phi$}&   \colhead{$ |\langle\vartheta\rangle - \phi|$}  &\colhead{$\mu$}  & \colhead{$\beta_{app}$} & && \colhead{$\alpha_m$}& \colhead{$\delta_m$}    \\  
\colhead{Source} & \colhead {I.D.} &  \colhead {N} & 
\colhead{(mJy)} &\colhead{(mas)} & \colhead{(pc)} & \colhead{(deg)}   & 
\colhead{(deg)}& \colhead{(deg)} &\colhead{($\mu$as y$^{-1})$}& \colhead{($c$)}  &\colhead{$t_{ej}$}  & \colhead{$t_\mathrm{mid}$}& \colhead{($\mu$as)}& \colhead{($\mu$as)}  \\  
\colhead{(1)} & \colhead{(2)} & \colhead{(3)} & \colhead{(4)} &  
\colhead{(5)} & \colhead{(6)} & \colhead{(7)} & \colhead{(8)} & 
 \colhead{(9)}& \colhead{(10)}&  
\colhead{(11)} & \colhead{(12)} & \colhead{(13)} & \colhead{(14)} & \colhead{(15)}   } 
\startdata 
0003+380  & 1 & 8  & 5 &4.23&  15.36& $ 120.7$ & 96$\pm$17 & 24$\pm$17 & 158$\pm$43 & 2.30$\pm$0.63 & \n &  2008.81 &3691$\pm$74 & $-$2169$\pm$80\\ 
0003+380  & 2 & 6  & 19 &1.78&  6.45& $ 112.6$ & 120.1$\pm$3.1 & 7.5$\pm$3.1 & 317$\pm$25 & 4.61$\pm$0.36 & \n &  2007.71 &1662$\pm$29 & $-$694$\pm$11\\ 
0003+380  & 4 & 5  & 16 &1.25&  4.53& $ 114.9$ & 205$\pm$14 & 90$\pm$14\tablenotemark{b} & 39$\pm$10 & 0.57$\pm$0.15 & \n &  2009.54 &1130$\pm$11 & $-$527$\pm$14\\ 
0003+380  & 5 & 8  & 40 &0.75&  2.71& $ 117.5$ & 21$\pm$89 & 96$\pm$89 & 2.7$\pm$7.6 & 0.04$\pm$0.11 & \n &  2010.26 &663$\pm$20 & $-$342$\pm$10\\ 
0003+380  & 6 & 10  & 98 &0.39&  1.43& $ 115.4$ & 335$\pm$46 & 141$\pm$46 & 12.7$\pm$8.4\tablenotemark{d} & 0.19$\pm$0.12 & \n &  2009.90 &350$\pm$22 & $-$158$\pm$19\\ 
0003$-$066  & 2 & 5  & 222 &1.05&  5.12& $ 322.9$ & 226.3$\pm$4.9 & 96.6$\pm$5.0\tablenotemark{b} & 191$\pm$15 & 4.09$\pm$0.33 & \n &  1997.80 &$-$585.9$\pm$8.9 & 883$\pm$37\\ 
0003$-$066  & 3 & 9  & 119 &2.82&  13.73& $ 296.9$ & 284.8$\pm$4.7 & 12.1$\pm$4.8 & 250$\pm$39 & 5.36$\pm$0.83 & \n &  1999.33 &$-$2375$\pm$98 & 1237$\pm$41\\ 
0003$-$066  & 4\tablenotemark{a} & 26  & 120 &6.61&  32.23& $ 285.6$ & 284$\pm$11 & 2$\pm$11 & 41$\pm$14 & 0.87$\pm$0.29 & \n &  2004.83 &$-$6326$\pm$60 & 1768$\pm$22\\ 
0003$-$066  & 5\tablenotemark{a} & 14  & 1031 &0.70&  3.40& $ 10.7$ & 350.9$\pm$5.3 & 19.9$\pm$5.5\tablenotemark{b} & 88.1$\pm$4.3 & 1.888$\pm$0.091 & \n &  2004.37 &138$\pm$18 & 634.1$\pm$9.0\\ 
0003$-$066  & 6\tablenotemark{a} & 10  & 97 &1.01&  4.92& $ 290.2$ & 210$\pm$15 & 81$\pm$15\tablenotemark{b} & 55$\pm$17 & 1.18$\pm$0.37 & \n &  2003.78 &$-$941$\pm$15 & 359$\pm$33
\enddata

\tablenotetext{a}{Acceleration model fit indicates significant accelerated motion.}
\tablenotetext{b}{Feature has significant non-radial motion according to the vector motion fit.}
\tablenotetext{c}{Feature has significant inward motion according to the vector motion fit.}
\tablenotetext{d}{Feature has slow pattern speed.}

\tablenotetext{~}{A question mark indicates a feature whose motion is not consistent with outward, radial motion but for which the possibility of inward motion and its degree of non-radialness are uncertain.}

\tablecomments{Columns are as follows: (1) B1950 name, (2) feature number, (3) number of fitted epochs, (4) mean flux density at 15 GHz in mJy,  (5) mean distance from core feature in mas, (6) mean projected distance from core feature in pc, (7) mean position angle with respect to the core feature in degrees, (8) position angle of velocity vector in degrees, (9) offset between mean position angle and velocity vector position angle in degrees,  (10) proper motion in $\mu$as y$^{-1}$, (11) apparent speed in units of the speed of light, (12) estimated epoch of origin, (13) date of reference (middle) epoch used for fit,  (14) fitted right ascension position with respect to the core at the middle epoch in $\mu$as, (15) fitted declination  position with respect to the core at the middle epoch in $\mu$as.}

\tablenotetext{}{(This table is available in its entirety in machine-readable form.)}

\end{deluxetable*}

A total of 881 of the robust features met the $\ge 10$ epoch criterion
for an acceleration fit, and we tabulate these results in
Table~\ref{accelmotiontablestub}.  The majority  (59\%) of these
well-sampled features display either significant acceleration or
non-radial motion, which confirms our previous finding that
accelerated motions are common in parsec scale AGN jets
\citep{MOJAVE_XIII}.

\begin{deluxetable*}{lcrrrrrrrrrr} 
\tablecolumns{12} 
%\rotate 
%\tabletypesize{\scriptsize} 
\tablewidth{0pt}  
\tablecaption{\label{accelmotiontablestub}Acceleration Fit Properties of Jet Features}  
\tablehead{\colhead{} & \colhead {}& \colhead{$\phi$} &    \colhead{$ |\langle\vartheta\rangle - \phi|$}& 
\colhead{$\mu$} & \colhead{$\beta_{app}$} & \colhead{$ \dot{\mu} $}  & \colhead{$ \psi $}  & \colhead{$ \dot{\mu}_\perp $}  & \colhead{$ \dot{\mu}_\parallel $}  &  \colhead{$\alpha_\mathrm{m}$} &\colhead{$\delta_\mathrm{m}$} \\  
\colhead{Source} & \colhead {I.D.} & \colhead{(deg)} & \colhead{(deg)} &   
\colhead{($\mu$as y$^{-1})$}& \colhead{(c)} &\colhead{($\mu$as y$^{-2})$}&\colhead{(deg)}& \colhead{($\mu$as y$^{-2})$}& \colhead{($\mu$as y$^{-2})$}  & \colhead{($\mu$as)}& \colhead{($\mu$as)}   \\  
\colhead{(1)} & \colhead{(2)} & \colhead{(3)} & \colhead{(4)} &  
\colhead{(5)} & \colhead{(6)} & \colhead{(7)}  & \colhead{(8)} & \colhead{(9)}  & \colhead{(10)}& \colhead{(11)}  & \colhead{(12)}  } 
\startdata 
0003+380  & 6  & $ 333 \pm 44 $ & $ 142 \pm 44 $& $ 13.4 \pm 8.6 $ & 0.20$\pm$0.12 &  $ 9.8 \pm 8.4$ & $ 309 \pm 53$ & $ -4.0 \pm 9.4$ & $ 9.0 \pm 9.0$     & $371\pm 33$ & $ -175 \pm 28$\\ 
0003-066  & 4\tablenotemark{a}  & $ 277.3 \pm 3.8 $ & $ 8.3 \pm 3.8 $& $ 50.9 \pm 5.3 $ & 1.09$\pm$0.11 &  $ 28.5 \pm 2.3$ & $ 73.7 \pm 3.1$ & $ 11.4 \pm 2.1$ & $ -26.1 \pm 2.5$     & $-6582\pm 32$ & $ 1693 \pm 20$\\ 
0003-066  & 5\tablenotemark{a}  & $ 353.9 \pm 3.0 $ & $ 16.8 \pm 3.1\tablenotemark{b} $& $ 87.2 \pm 4.4 $ & 1.868$\pm$0.093 &  $ 26.6 \pm 4.9$ & $ 274 \pm 10$ & $ -26.3 \pm 4.9$ & $ 4.5 \pm 4.8$     & $199\pm 15$ & $ 630 \pm 14$\\ 
0003-066  & 6\tablenotemark{a}  & $ 211.3 \pm 9.6 $ & $ 78.9 \pm 9.6\tablenotemark{b} $& $ 54 \pm 11 $ & 1.16$\pm$0.24 &  $ 65 \pm 16$ & $ 336 \pm 11$ & $ 54 \pm 13$ & $ -37 \pm 18$     & $-901\pm 16$ & $ 268 \pm 35$\\ 
0003-066  & 8\tablenotemark{a}  & $ 290.7 \pm 1.6 $ & $ 3.5 \pm 1.6 $& $ 330.4 \pm 9.7 $ & 7.08$\pm$0.21 &  $ 67 \pm 12$ & $ 127 \pm 10$ & $ -19 \pm 12$ & $ -64 \pm 12$     & $-2444\pm 30$ & $ 1121 \pm 28$\\ 
0003-066  & 9  & $ 295.2 \pm 4.1 $ & $ 7.5 \pm 4.3 $& $ 278 \pm 20 $ & 5.96$\pm$0.42 &  $ 99 \pm 35$ & $ 110 \pm 22$ & $ 9 \pm 37$ & $ -99 \pm 35$     & $-1769\pm 52$ & $ 582 \pm 53$\\ 
0010+405  & 1  & $ 340.7 \pm 4.4 $ & $ 11.9 \pm 4.4 $& $ 432 \pm 42 $ & 6.99$\pm$0.68 &  $ 44 \pm 83$ & $ 147 \pm 76$ & $ 11 \pm 53$ & $ -43 \pm 70$     & $-4259\pm 76$ & $ 6991 \pm 107$\\ 
0010+405  & 2  & $ 9 \pm 123 $ & $ 41 \pm 123 $& $ 2 \pm 14 $ & 0.04$\pm$0.23 &  $ 4 \pm 22$ & $ 152 \pm 123$ & $ 2 \pm 21$ & $ -3 \pm 23$     & $-898\pm 30$ & $ 1470 \pm 48$\\ 
0010+405  & 3  & $ 138 \pm 83 $ & $ 170 \pm 83 $& $ 2.5 \pm 5.4 $ & 0.041$\pm$0.088 &  $ 6.9 \pm 6.1$ & $ 99 \pm 57$ & $ -4.3 \pm 8.8$ & $ 5.4 \pm 9.0$     & $-493.6\pm 9.5$ & $ 783 \pm 15$\\ 
0010+405  & 4  & $ 113 \pm 98 $ & $ 145 \pm 98 $& $ 1.4 \pm 4.5 $ & 0.022$\pm$0.072 &  $ 5.2 \pm 8.9$ & $ 318 \pm 69$ & $ -2.2 \pm 7.1$ & $ -4.7 \pm 8.1$     & $-240.6\pm 9.0$ & $ 382 \pm 14$
\enddata 
\tablenotetext{a}{Feature shows significant accelerated motion.}
\tablenotetext{b}{Feature shows significant non-radial motion according to the acceleration fit.}
\tablenotetext{c}{Feature shows significant inward motion according to the acceleration fit.}
\tablenotetext{~}{A question mark indicates a feature whose motion is not consistent with outward, radial motion but for which the possibility of inward motion and its degree of non-radialness are uncertain.}

\tablecomments{Columns are as follows: (1) B1950 name, (2) feature number, (3) proper motion position angle in degrees,  (4) offset between mean position angle and proper motion position angle in degrees, (5) proper motion in $\mu$as  y$^{-1}$, (6) apparent speed in units of the speed of light, (7) acceleration in $\mu$as  y$^{-2}$, (8) acceleration vector position angle in degrees, (9)  acceleration perpendicular to velocity direction in $\mu$as  y$^{-2}$, (10)  acceleration parallel to velocity direction in $\mu$as  y$^{-2}$, (11) fitted right ascension position with respect to the core at the middle epoch in $\mu$as, (12) fitted declination  position with respect to the core at the middle epoch in $\mu$as. }

\tablenotetext{}{(This table is available in its entirety in machine-readable form.)}

\end{deluxetable*}

\begin{figure*}
\begin{center}
\includegraphics[angle=270,width=0.95\textwidth]{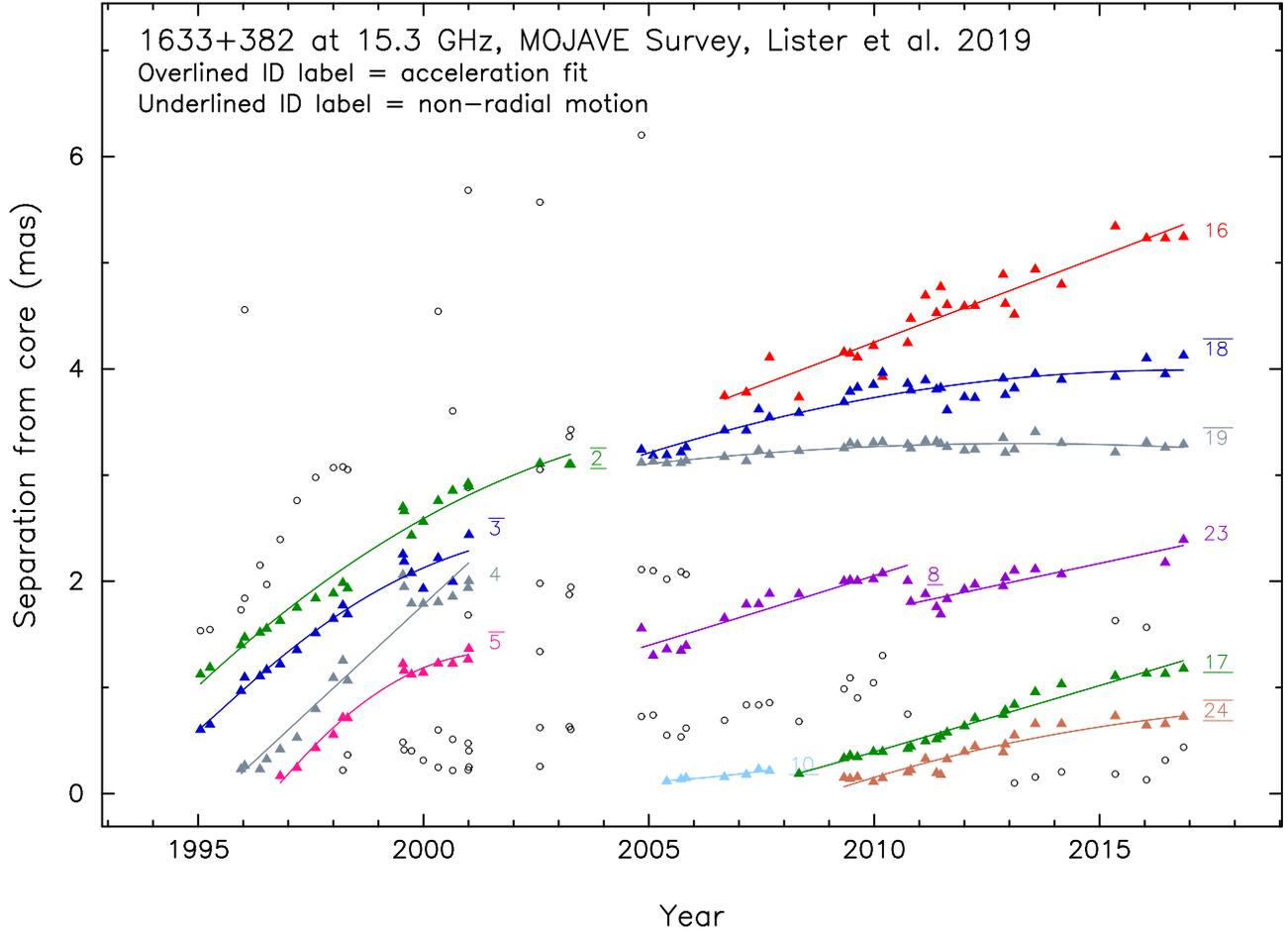}
\end{center}
\caption{\label{sepvstime} Plot of angular separation
  from the core versus time for Gaussian jet features. The B1950
  source name is given at the top left of each panel. Colored symbols
  indicate robust features for which kinematic fits were obtained. The
  identification number is overlined if the acceleration model was fit
  and indicated a $>3\sigma$ acceleration. An underlined
  identification number indicates a feature with non-radial motion.
  The $1 \sigma$ positional errors on the individual points typically
  range from 10\% of the FWHM restoring beam dimension for isolated
  compact features, to 20\% of the FWHM for weak features. This
  corresponds to roughly 0.03 mas to 0.15 mas, depending on the source
  declination.  (This is a figure stub; an extended version is
  available online.) }
\end{figure*}

\begin{figure*}
\begin{center}
\includegraphics[angle=270,width=0.95\textwidth]{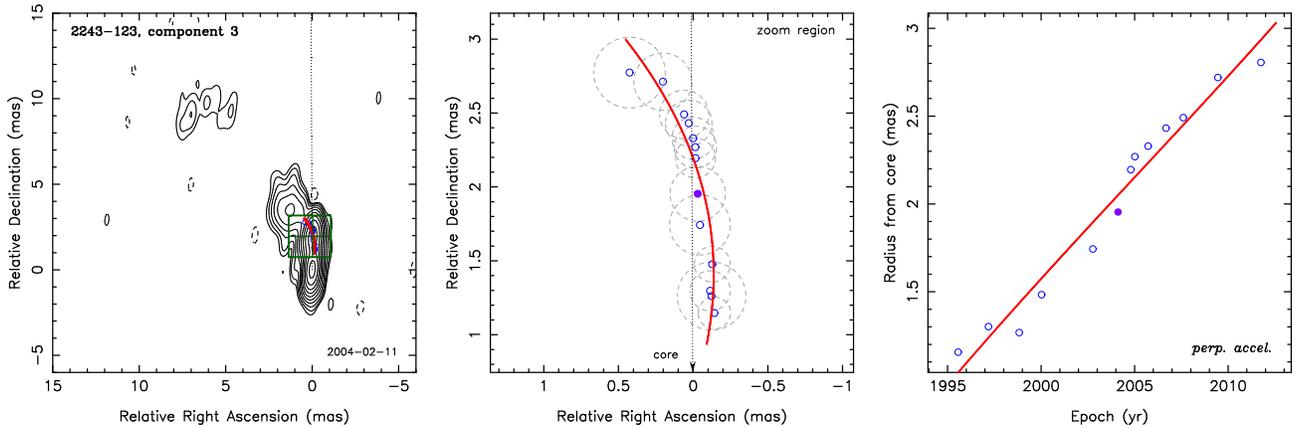}
\end{center}

\caption{\label{xyplot} Motion fits and sky position plots
  of individual robust jet features. Positions are
  relative to the core position.  The left-hand panel shows a 15 GHz
  VLBA total intensity contour image of the jet at the epoch closest to the middle
  reference epoch. The green box delimits the zoomed region that is
  displayed in the middle panel. The feature's position at the
  image epoch is indicated by the green cross-hairs. The dotted line
  connects the feature with the core feature and is plotted with the
  mean position angle.  The position at the image epoch is shown
  by a filled blue circle while other epochs are plotted with
  unfilled blue circles. The red solid line indicates the vector fit
  (or accelerating fit, if there is significant acceleration) to the
  feature positions. The gray dashed circles / ellipses indicate the
  fitted FWHM sizes of the feature at the measured epochs.  (This is a figure stub; an extended
  version is available online.) }
\end{figure*}

In Figure Set~\ref{sepvstime} we plot the angular separation of
features from the core in each jet versus time. The robust features
are plotted with filled colored symbols and solid lines representing
the fit. The feature identification number is overlined if
the acceleration model was fit and yielded a $>3\sigma$ acceleration.
An underlined identification number indicates a feature with
non-radial motion, i.e., its velocity vector did not point back to the
core location within the errors.  We plot the individual trajectories
and fits on the sky for all the robust features in Figure
Set~\ref{xyplot}.

\subsubsection{Pattern Speeds}

In a previous kinematic study \citep{MOJAVE_X}, we found that in many
individual AGN jets, there is no single apparent speed
$\beta_\mathrm{app} = v_\mathrm{app}/c$ at which bright features
propagate downstream. Instead, there is typically a single
characteristic speed with a modest spread around this value.  Since
trackable features emerge only every few years in most bright blazar
jets, continuous monitoring periods of a decade or more are needed to
establish the characteristic speed of a jet, and whether any
individual feature may have an atypically low pattern speed (see also
a recent analysis of MOJAVE kinematics results by \citealt{2018arXiv181102544P}).

\begin{figure}
\begin{center}
\includegraphics[angle=0,trim=0cm 0cm 0cm 0cm,clip,width=0.95\columnwidth]{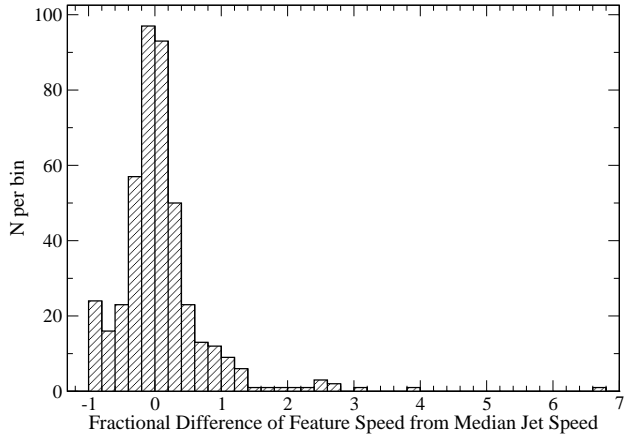}
\end{center}
\caption{\label{speed_dispersion} Overall normalized speed
  distribution within jets with at least ten robust features. The
  fractional difference is defined as $(\mu - \mu_\mathrm{median}) /
  \mu_\mathrm{median}$. }
\end{figure}

In Figure~\ref{speed_dispersion} we show the distribution of speed
differences from the jet's median speed for 436 features in 26 jets
that have ten or more robust features.  This plot contains nearly
twice as many jet features as our previous kinematic study, and is qualitatively
similar.  Most features lie within $\pm 40$\% of the jet's median
speed. There is also a small tail consisting of atypically fast
features.  The jet with the largest range of speeds is 4C $+$15.05
(0202$+$149), which has ten features with apparent speeds ranging
from 0.1 $c$ to 16 $c$.

\begin{figure}
\begin{center}
\includegraphics[angle=0,trim=0cm 0cm 0cm 0cm,clip,width=0.95\columnwidth]{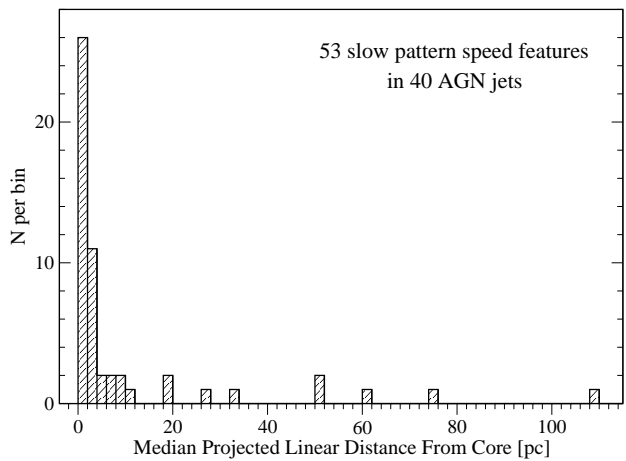}
\end{center}
\caption{\label{spsdist_pc} Distribution of projected
  linear distance from the core feature in parsecs for 53 features
  classified as having a slow pattern speed.}
\end{figure}

We have identified 55 features in 42 AGN jets that have appreciably
slower speeds than other features in the same jet. Our specific
criteria are that the feature (i) does not have a $>3\sigma$
acceleration, (ii) has an angular speed smaller than 20 \muasyr, and
(iii) has a speed at least ten times slower than the fastest feature
in the same jet.  Figure~\ref{spsdist_pc} shows the distribution of
projected distance from the core for 53 slow pattern speed features in
40 AGNs with known redshifts. The vast majority are located within 4
pc of the core feature ($\sim 100$ pc de-projected, given typical
viewing angles $< 2^\circ$). This is consistent with the 43 GHz VLBA
survey of 36 AGNs by \cite{2017ApJ...846...98J}, who found 21\% of jet
features to be quasi-stationary, with most located at projected core
distances below 3 pc.

Of the 1744 robust jet features that we have studied, only 44 (2.5\%) have 
`velocity vectors that are directed inward toward the core feature. We
might expect to see rare instances of apparent inward motion when a feature
moving along a curved trajectory crosses our line of sight (e.g., as
in the case of 4C +39.25; \citealt{2000AA...361..529A}). It is also
possible that small changes in the brightness distribution of a large
diffuse feature may alter its best-fit Gaussian centroid location,
creating apparent inward motion. We note two instances (feature id = 1
in 87GB 061258.1+570222 and id = 1 in 8C 1944+838) where this may be the
case. Inward motion can also result from incorrect identification of
the core feature, or variable structure near the core that is below
the angular resolution of our observations that may alter the fitted
core location. We note that in 16 of 33 AGN jets with inward motion,
the inward-moving feature is the closest feature to the core, and four of
these jets (all associated with BL Lac objects: UGC 00773, 3C 66A, Mrk
421, ON 325) have more than one close-in inward-moving feature.

\subsubsection{\label{speeddistrib}Speed Distributions}

\begin{figure}
\begin{center}
\includegraphics[angle=0,width=0.95\columnwidth]{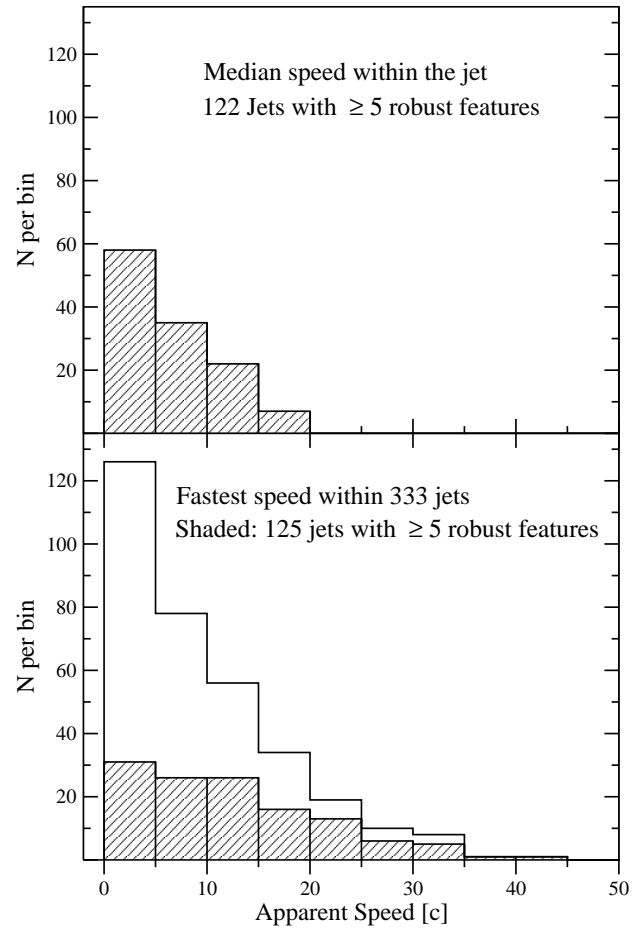}
\end{center}
\caption{\label{betaspeedhist} Top panel: distribution of
  median apparent speed within 122 AGNs having at least five robust jet
  features. Bottom panel: distributions of maximum apparent speed for
  333 AGN jets (unshaded) and 125 AGNs having at least five robust jet
features (shaded). }
\end{figure}

We have calculated maximum and median speed statistics for the jets in
our sample using the method described in \cite{MOJAVE_X}. For
accelerating features, we note that the speeds are determined at the
middle epoch, and thus may not represent the maximum speed attained by
the feature.  In the case of two AGNs for which we could not identify
any robust features (AO 0235+164 and 1ES 1959+650), we adopted maximum
speeds from the literature based on VLBA observations made at other
wavelengths. We plot the distributions of these statistics in
Figure~\ref{betaspeedhist}. Slow apparent speeds are common, with very
few measured speeds above 30 $c$. As discussed by \cite{VC94} and
\cite{LM97}, the shape of the distributions is incompatible with all
jets having the same bulk Lorentz factor, and instead suggests a power
law parent distribution that is weighted towards slow speeds.
Single-valued parent $\Gamma$ distributions predict an excess of high
apparent jet speeds, while Gaussian $\Gamma$ distributions do not
reproduce the gradual fall-off in the number of jets with higher
apparent speeds.

\subsubsection{Statistical Trends}

\begin{figure*}
\begin{center}
\includegraphics[angle=270,width=0.95\textwidth]{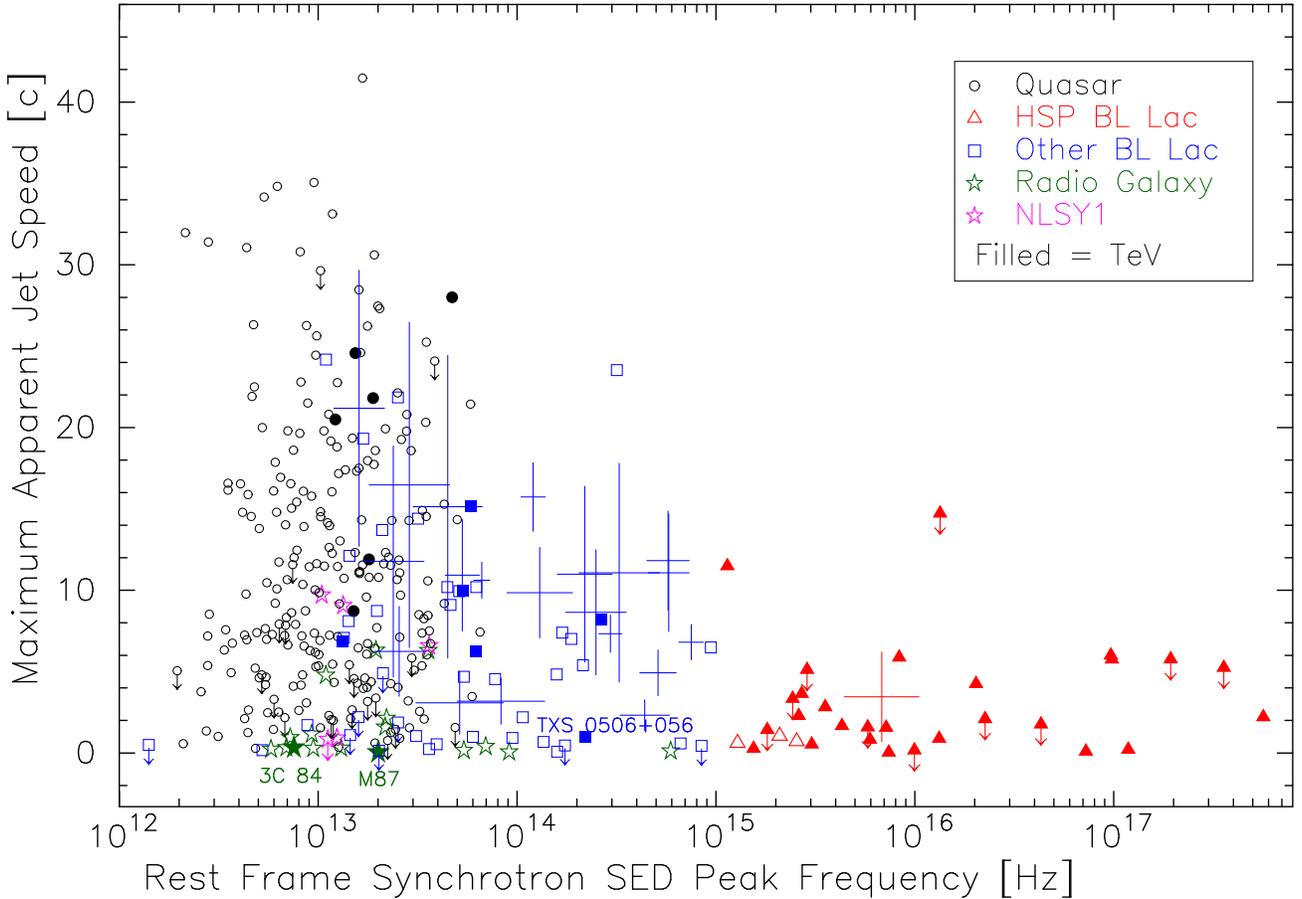}
\end{center}
\caption{\label{beta_max_vs_sed_peak} Maximum apparent jet
  speed versus synchrotron peak frequency for jets in the MOJAVE
  survey, as well as those in the survey of
  \cite{2018ApJ...853...68P}. Upper limit values are denoted by
  downward arrows. Quasars are indicated by black circles, radio
  galaxies by green stars, narrow-line Seyfert 1 galaxies by violet
  stars, high synchrotron peaked BL Lac objects by red triangles, and
  other BL Lac objects by blue squares. Filled symbols indicate
  detections by ground-based TeV gamma-ray observatories. The cross
  symbols indicate BL Lacs for which only upper and lower limits on
  the redshift are known. }

% Produced with maxbeta_plot.pl

\end{figure*}

In Figure~\ref{beta_max_vs_sed_peak} we plot maximum apparent jet
speed versus rest frame synchrotron SED peak frequency. The plot
includes AGNs from our survey, as well as those from
\cite{2018ApJ...853...68P} and \cite{2017ApJ...846...98J}.  AGNs with $<3
\sigma$ maximum speeds are indicated with upper limit symbols. The
crosses indicate BL Lacs with no known redshift, and their extents
correspond to lower and upper redshift limits published in the
literature. For clarity, we have omitted BL Lacs for which the redshift
limits give a possible range of $\beta_\mathrm{app}$ greater than 20.
There is a clear upper envelope to the distribution, with the highest
jet speeds being found only in AGNs with low synchrotron peak
frequencies.

The filled symbols indicate AGNs that have been detected at TeV
gamma-ray energies with the airshower telescopes VERITAS, HESS, or
MAGIC. The large fraction of high synchrotron peaked (HSP) AGNs that
are TeV-detected in this plot is a selection effect since these have
been specifically targeted for long term VLBA kinematic study by
\cite{2018ApJ...853...68P}. The ISP AGNs have been targeted in MOJAVE
on the basis of their detection at GeV energies by \Fermi, while most
of the low synchrotron peaked (LSP) AGNs are from the radio-selected
MOJAVE sample \citep{MOJAVE_XIII}.  Although a fast jet speed does not
guarantee a TeV detection, it does appear to be a minimum requirement
for the intermediate- and low synchrotron peaked AGNs. This implies a
direct connection between the bulk jet speed measured on parsec scales
and the Doppler boosting level of the TeV emission. Of the 14 non-HSP
TeV detected AGNs in Figure~\ref{beta_max_vs_sed_peak}, only three
have maximum jet speeds below 6 c. Two of these (3C 84 and M 87)
are very nearby ($< 75$ Mpc) radio galaxies, and the third (TXS
0506+056) is an unusual ISP BL Lac with a measured maximum speed of
$0.98 c \pm 0.3 c$ that lies within the sky error circle of a
high-energy neutrino event detected in 2017 \citep{2018Sci...361..147I}.

\section{MONTE CARLO JET PARENT POPULATION MODELING}
\label{montecarlo}

The interpretation of parsec scale AGN jet kinematic studies presents
a challenge in the sense that the individual objects that are most
easily studied (i.e., high flux density, with proper motions
observable on time periods of $\sim$ a few years) are blazars, whose
selection is highly affected by Doppler bias
\citep{1979Natur.277..182S}.  In principle, the observed redshift,
luminosity, and apparent speed distributions of a complete flux
density-limited jet sample can be used to recover the intrinsic
properties of the blazar parent population, but the Doppler selection
effects need to be carefully accounted for.  \cite{VC94} and \cite{LM97}
have shown that this can be done analytically only in the case of very
simplistic, non-realistic assumptions.  These include a non-evolving
single power-law luminosity function and a single-valued or uniform
distribution of bulk Lorentz factors, neither of which provide
satisfactory fits to the data. The typical approach (e.g.,
\citealt{LM97,MOJAVE_VI,2008AJ....136.1533B, 2012MNRAS.420.2899G,
  2015MNRAS.451.2434L}) has been to generate simulated flux
density-limited samples from jet parent populations whose properties
are drawn from specified probability distributions, and find the set
of distribution parameters that best fit the data.  In this section we
carry out this type of Monte Carlo analysis on our MOJAVE data, based
on the method of \cite{LM97}.

\subsection{Simulated Jet Properties}

The observed flux density $S_\nu$ from a spherical optically thick
source of radiation with an isotropically emitted rest frame
luminosity $L_\nu \propto \nu^\alpha$, moving with bulk Lorentz factor
$\Gamma$ at an angle $\theta$ to the line of sight and located at
redshift $z$ (with corresponding luminosity distance $D_L(z)$) is
(e.g., \citealt{BlandfordKonigl79, 2018PASP..130g3001C})
\begin{equation}\label{fluxequation}
S_\nu = { L_\nu \delta^p \; (1+z)^{(1+\alpha)}  \over 4\pi D_L^2(z)},
\end{equation}

where $\nu$ is the observing frequency and $L_\nu$ is the luminosity
emitted in the jet frame at that same frequency.  The exponent $p$ of
the Doppler factor
 
\begin{equation}
\delta = \left[\Gamma -  \sqrt{(\Gamma^2-1)}\cos{\theta}\right]^{-1}
\end{equation}

is $p = 3 - \alpha$ in the scenario described above.  However, in the case of a
continuous jet made up of many such spheres, one cannot distinguish
the lifetimes of the individual emitting particles, and a time
dilation factor of $\delta$ is no longer applicable, hence $p = 2 -
\alpha$ \citep{Cawthorne1991}.

The minimum properties required to simulate the observed flux density
of an AGN jet are therefore $z$, $L_\nu$, $\Gamma$, $\theta$,
$\alpha$, and $p$.  Actual AGN jets present complications in terms of
the geometry of their emitting regions, optical depth variations, and
flow accelerations, but the highest contributions to the observed flux
density will come from regions where the synchrotron emission coefficient is
highest, and where the Doppler factor is largest (e.g., $\theta
\lesssim \cos^{-1}{\beta}$, where $\beta$ is the flow velocity in
units of the speed of light).  Stacked-epoch MOJAVE VLBA images of
blazars show mainly conical jet profiles \citep{MOJAVE_XIV} in which
adiabatic expansion and synchrotron losses exponentially reduce the
electron energies and magnetic field strength with distance down the
jet (e.g., \citealt{1981ApJ...243..700K}). The bulk of the synchrotron
emission therefore originates near the base of the jet, as confirmed
by VLBI morphologies that typically consist of a bright optically
thick core feature accompanied by a much weaker jet.  The exceptions
to this are (i) young AGN jets of the CSO/GPS class, which have high
luminosity radio lobes that are interacting with the interstellar
medium of the host galaxy \citep{1998PASP..110..493O}, and (ii) rare
instances where a bent downstream jet flow crosses the line of sight
and experiences maximum Doppler boosting (e.g., 4C +39.35,
\citealt{2000AA...361..529A}).

There are therefore good reasons to expect that a simulated population
where each jet consists of a single (core) emitting region can provide
a good representation of a suitably chosen blazar sample. The 1.5JyQC
sample is well-suited in several respects, as it is a complete flux
density-limited sample selected at high radio frequency, where the
relative flux density contribution of the steep-spectrum downstream jet
emission is low compared to the (typically flat-spectrum) core. It is
also selected on the basis of VLBI flux density, which includes no
contribution from any large kiloparsec scale emission. Any
contaminating CSO/GPS sources can be rejected on the basis of
available spectral and morphological information, and most
importantly, the sample is large enough to statistically constrain the
best fit parameters of the Monte Carlo simulations.  After dropping
two GPS quasars (PKS B0742+103 and OI $-$072) and
six AGNs with no optical spectral information there are 174 1.5JyQC
quasars with redshift $z \ge 0.15$ suitable for comparison with our
simulations.

\subsection{Simulation Parameters}

Our simulation method is to generate a parent population of jets drawn
from specified redshift, Lorentz factor, radio luminosity, and viewing angle
distributions, calculate their predicted flux densities, and retain
those jets that exceed the specified 1.5 Jy flux density limit.  Because
the 1.5JyQC sample includes all AGNs above declination $-30^\circ$
known to have exceeded 1.5 Jy over a 25 year period, we do not include
any flux variability in our simulations, but instead compare our
simulated jet flux densities to the maximum jet flux density for each
AGN measured during the 1.5JyQC selection period (column 5 of
Table~\ref{gentable1.5}).

\subsubsection{Luminosity Function}

Despite many studies on the radio luminosity functions (LFs) of AGNs,
there is still no consensus on whether radio-loud AGN LFs evolve with
lookback time in a manner consistent with increasing number density,
increasing luminosity, or a mixture of both
\citep{2014MNRAS.445..955B,2017AA...602A...6S,2018ApJS..239...33Y}.
There are also indications that lower power (i.e., FR I) AGNs may
evolve differently than the high power (FR II) population
\citep{2008MNRAS.385..310R}.  Given these uncertainties, we have
adopted a simple pure luminosity evolution parameterization for flat
spectrum radio quasars used by \cite{2012ApJ...751..108A} and
\cite{2017ApJ...842...87M}:
\begin{equation}
  \Phi(L,z) \propto \Phi(L/e(z)),
\end{equation}
where 
\begin{equation}
  e(z) = (1+z)^{k} e^{z/\eta},
\end{equation}
and
\begin{equation}
  \Phi(L/e(z=0)) \propto L^\gamma.
\end{equation}

Our approach is to find the best fit values of $\gamma$, $\eta$ and
$k$ using the MOJAVE data.  We restrict our comparisons to quasars in
the 1.5JyQC sample only, given the possibility that the BL Lac objects may be
drawn from a different (i.e., lower power, or FR I) parent population
\citep{UP95}. We set the lower limit on the parent LF at $10^{24}$ W
Hz$^{-1}$ based on the least powerful known FR II radio galaxies
(e.g., \citealt{2012ApJ...756..116A}).

\subsubsection{Redshift Distribution}

By adopting a pure luminosity evolution model, we assume that the
parent jet population has a constant co-moving density with redshift.
All of the 1.5JyQC quasars have redshifts greater than 0.15, with the
exception of TXS 0241+622 ($z = 0.045$). In order to avoid small
number statistics in this nearby volume of space, we drop this AGN
from our data comparisons and set the lower redshift limit of our
simulation to $z = 0.15$. Because the form of LF evolution is not well
known at very high redshift, we set the upper redshift limit in our
simulations to that of the highest redshift 1.5JyQC quasar: OH 471 ($z
= 3.4$).

\begin{figure*}
\begin{center}
\includegraphics[angle=0,trim=2cm 1cm 2cm 2cm,clip,width=0.95\textwidth]{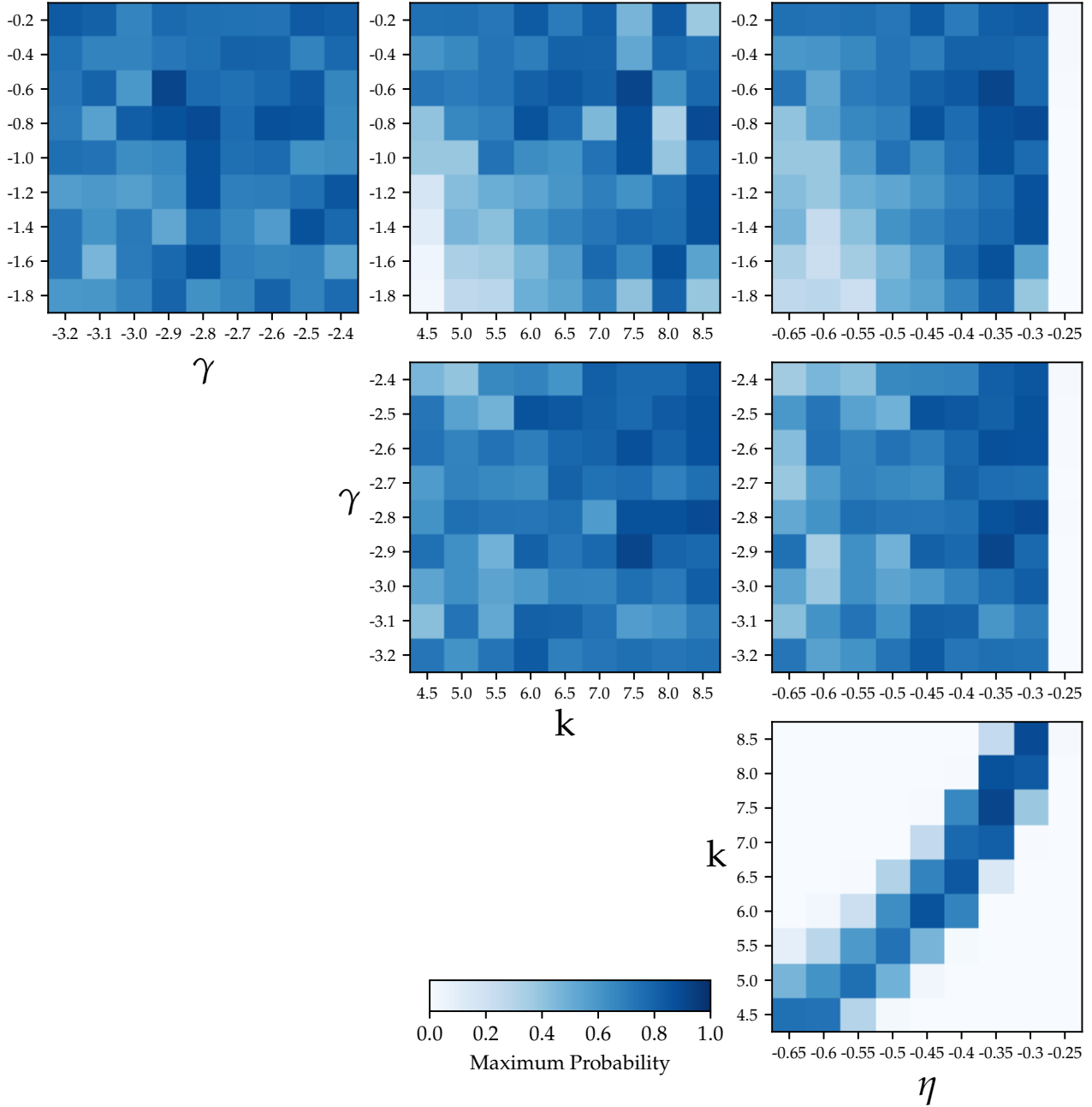}
\end{center}
\caption{\label{p_z} Corner plot showing two-dimensional
  parameter space projections of the Lorentz factor distribution power
  law index $b$ and luminosity function evolution parameters $k$ and
  $\eta$ for the Monte Carlo parent population simulation grid with
  Doppler boosting index $p = 2$. The false color scale corresponds to
  the maximum A-D test probability that the redshifts of
  the 1.5JyQC quasar sample and a simulation having that particular
  parameter combination are drawn from the same parent population.
  Lighter colors indicate poorer fits to the data. }
  
% Created with colorplot_AD_tests.py
\end{figure*}

\begin{figure*}
\begin{center}
\includegraphics[angle=0,trim=2cm 1cm 2cm 2cm,clip,width=0.95\textwidth]{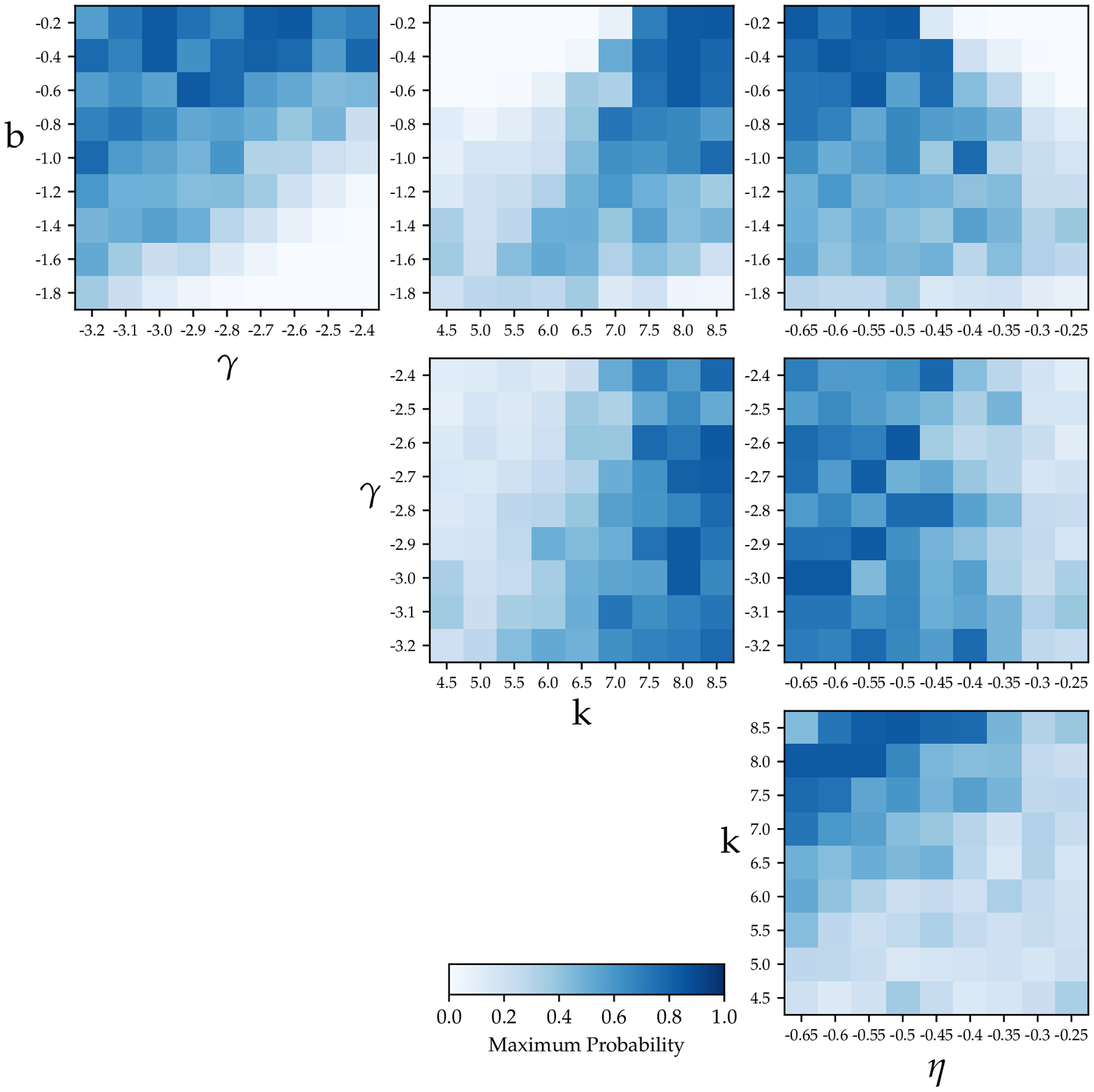}
\end{center}
\caption{\label{p_bapp}
Corner plot showing two-dimensional
  parameter space projections of the Lorentz factor distribution power
  law index $b$ and luminosity function evolution parameters $k$ and
  $\eta$ for the Monte Carlo parent population simulation grid with
  Doppler boosting index $p = 2$. The false color
  scale corresponds to the maximum A-D test probability that the
  apparent jet speeds of the 1.5JyQC quasar sample and a simulation
  having that particular parameter combination are drawn from the same
  parent population. Lighter colors indicate poorer fits to the
  data. }
% Created with colorplot_AD_tests.py
\end{figure*}

\begin{figure*}
\begin{center}
\includegraphics[angle=0,trim=2cm 1cm 2cm 2cm,clip,width=0.95\textwidth]{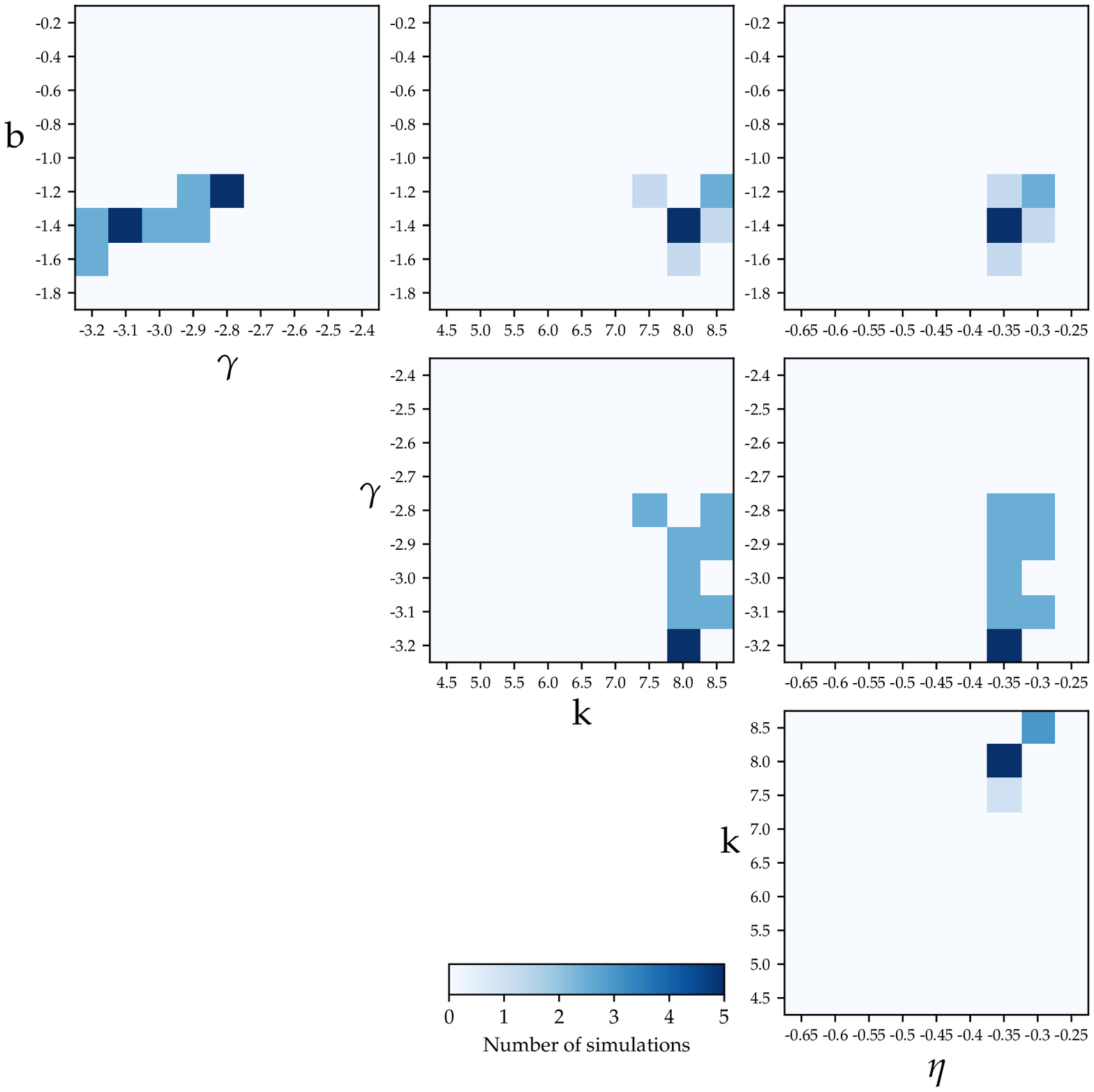}
\end{center}
\caption{\label{acceptable_fits} Corner plot showing
  two-dimensional parameter space projections of the Lorentz factor
  distribution power law index $b$ and luminosity function evolution
  parameters $k$ and $\eta$ for the Monte Carlo parent population
  simulation grid with Doppler boosting index $p = 2$. The false color
  scale corresponds to the number of simulations having that
  particular parameter combination that provide acceptable fits to the
  1.5JyQC sample data.  }
% Created with colorplot_AD_tests.py
\end{figure*}

\subsubsection{Bulk Lorentz Factor Distribution and Doppler Boosting
  Index}

Due to the strong selection biases associated with Doppler boosting,
any large flux density-limited jet sample should contain some jets
with the maximum Lorentz factor in the population (viewed at small
$\theta$). In the MOJAVE sample the fastest instantaneous measured jet
speed is approximately 50 $c$ for an accelerating feature in the jet
of PKS 0805$-$07 \citep{MOJAVE_XIII}, which corresponds to a
$\Gamma_\mathrm{max} \simeq 50$ . In light of our discussion
of the observed apparent velocity distributions in
\S~\ref{speeddistrib}, we adopt a power law Lorentz factor
distribution for our simulated jets of the form $N(\Gamma) \propto
\Gamma^b$, where $b$ is a free parameter with values less than zero and $\Gamma$
ranges from 1.25 to 50.  The lower limit on $\Gamma$ ($\beta \simeq
0.6 c$) is based on a Bayesian analysis of the relative prominence of
radio cores and kiloparsec scale jets in FR II radio sources by
\cite{2009MNRAS.398.1989M}. We assume no evolution of the jet Lorentz
factor distribution with redshift.

The brightest radio-loud AGN cores are known to have a range of
spectral indices with a mean value  $\alpha = 0.22 \pm 0.03$
\citep{MOJAVE_XI}, however, the intrinsic distribution is not well
known due to the difficulty of deconvolving relativistic beaming and
projection effects. The spectral index enters into the simulated jet
flux density via the small $(1+z)^{1+\alpha}$ k-correction, and more
importantly, the Doppler boost index $p$. Any spread of $\alpha$ in
the parent population will be effectively smoothed out in the observed
luminosity function, so we fix $\alpha = 0$ for all our simulated jets
and assume continuous jet emission such that $p = 2$. We discuss other
fixed values of $\alpha$ in \S~\ref{simcomparisons}.

The Monte Carlo analysis of \cite{LM97} included the possibility of an
intrinsic correlation between jet Lorentz factor and synchrotron
luminosity of the form $L \propto \Gamma^\xi$.  They found that both
$\xi = 0$ and $\xi \neq 0$ models produced very similar fits to the
Caltech-Jodrell Flat-Spectrum AGN sample data. As we will show in
Section~\ref{simcomparisons}, we are able to obtain good fits to the
1.5JyQC quasar sample assuming no $L-\Gamma$ correlation, so we
explore only the $\xi = 0$ case in this paper.

\subsection{Simulation Procedure}

In order to search for the best fit parent population parameters, we
constructed a grid of simulations with equally spaced parameter values
spanning the ranges listed in Table~\ref{MCparams}. The procedure used
to create each simulation in the grid is as follows:

(i) Select values for $b$, $\gamma$, $k$, and $\eta$. 

(ii) Generate $z$, $L_\nu$, $\Gamma$, and $\theta$ values for a single jet
from the probability distributions listed in Table~\ref{MCparams}.

(iii) Calculate the observed flux density of the jet
according to Equation~\ref{fluxequation}.  We ignore any 
contribution from the counter-jet since it will be negligible for AGNs
in a highly Doppler-biased sample (see \S~\ref{bestfitsim}).

(iv) If $S_\nu > 1.5$ Jy, keep the simulated jet.

(v) Repeat steps (ii) through (iv) until a sample of jets 10 times larger
than the 1.5JyQC comparison sample is obtained, and
record the total size of the parent population needed to produce this
sample.

\begin{deluxetable*}{llll}
\tablecolumns{4} 
\tablewidth{0pt}  
\tablecaption{\label{MCparams} Monte Carlo Jet Model Parameters}  
\tablehead{\colhead{Jet Property} & \colhead {Distribution} &  
\colhead{Fixed parameters} & \colhead{Free parameter ranges}}
\startdata 
Lorentz factor & $N(\Gamma)d\Gamma \propto \Gamma^b$ &
$\Gamma_\mathrm{min} = 1.25$ &  $-1.8 \le b\le -0.2$, step = 0.2 \\
&&  $\Gamma_\mathrm{max} = 50$ \\
\\

Luminosity function & $ \Phi(L,z) \propto \Phi(L/e(z)) $&
$L_\mathrm{min} = 10^{24}$ W Hz$^{-1}$ & $-0.65 \le \eta \le -0.25$, step = 0.05 \\

&$ e(z) = (1+z)^{k} e^{z/\eta} $  &    $L_\mathrm{max} = 10^{31}$ W
Hz$^{-1}$ &  $4.5\le k\le 8.5$, step = 0.5  \\

&$ \Phi(L/e(z=0)) \propto L^\gamma$ &&  $-3.2 \le \gamma \le -2.4$, step = 0.1\\

\\
Beamed luminosity & $P = L\, \delta^p $ & $p = 2 + \alpha$ & $\alpha = -0.5$, 0, 0.22 \\
\\
Viewing angle & $p(\theta)d\theta = \sin{\theta}$ & $\theta_\mathrm{min}
= 0^\circ$   & \n \\
&& $\theta_\mathrm{max} = 90^\circ$

\enddata 
\end{deluxetable*}

By creating larger samples than the data sample in step (v) we reduce
the amount of statistical fluctuations associated with selecting a
relatively small number of bright AGNs from a very large parent
population.  In doing so, we are effectively creating simulated jet
samples from ten Universes and are comparing the mean properties of
these samples to the data.

\subsection{\label{simcomparisons}Comparisons to MOJAVE Data}

For each simulation in the four dimensional parameter grid ($b$,
$\gamma$, $k$, $\eta$) we compared the simulated flux density, redshift, radio
luminosity ($P_\nu$), and apparent velocity distributions to the 1.5JyQC
sample of 174 quasars using the Anderson-Darling (A-D) test.  The
latter is a non-parametric test that assesses whether two samples are
drawn from different parent populations, and is sensitive to a wider
variety of possible distribution differences than the frequently used
Kolmogorov-Smirnov test \citep{Engmann2011}.  Our method was to
randomly select a sample of 174 jets from the simulation and perform
the A-D tests against the 1.5JyQC sample. We repeated this process 10 times
and recorded the median A-D test probabilities $p_S$, $p_z$, $p_P$, and
$p_{\beta\mathrm{app}}$ , corresponding to the probability of the null
hypothesis that the simulated and 1.5JyQC distributions are drawn from
the same parent population.  Since the completeness of the
observational data is high (100\% for $S$, $z$ and $P$, and 87\% for
$\beta_\mathrm{app}$), we did not use any bootstrapping procedures
to simulate the missing data.

In Figure~\ref{p_z} we plot two-dimensional projections of the grid
parameter space, where the false-color corresponds to the maximum
value of $p_z$ for any simulation having that particular parameter
combination. The 1.5JyQC redshift distribution serves to constrain the
parameter space to a limited number of $k-\eta$ (evolution parameter)
combinations, as seen in the lower right panel. The top row of plots
in Fig.~\ref{p_z} indicates, however, that the 1.5JyQC redshifts can
be well-reproduced with many different combinations of the parent LF
and the Lorentz factor distribution parameters.

The two-dimensional projections in Figure~\ref{p_bapp}, in which the
false-color corresponds to the maximum values of
$p_{\beta\mathrm{app}}$, serve to further constrain the region of
viable parameter space for the simulations.  The $\beta_\mathrm{app}$
distribution is best fit with simulations with $k > 5.5$.  Also, the
values of $k$ and $\eta$ that provide the best fits to the 1.5JyQC
apparent velocity distribution (lower right panel) yield relatively
poor fits to the observed luminosity distribution.

Within the full grid, the simulation with the highest A-D
probability summed over all four observable quantities has $b = -1.4$, $\gamma
= -3.1$, $k = 8.0$, $\eta=-0.35$ (model A). 
There are no other simulations in the grid that have an A-D
probability greater than 0.4 in all four quantities. We investigated
the effect of random statistical outliers on the A-D probability
values for this best fit simulation by first creating a simulated
flux density-limited sample of 174000 jets (i.e., 1000 Universes), then
selecting a random subset of 174 jets to compare with the 1.5JyQC
data. After repeating the random subset selection 1000 times, the
standard deviations on $p_S$, $p_z$, $p_P$, and
$p_{\beta\mathrm{app}}$ were 0.25, 0.3, 0.25, and 0.2 respectively. We
therefore consider any simulation that has all four A-D probabilities
within 1 $\sigma$ of those of the best fit simulation to also be an
acceptable fit to the data.

In Figure~\ref{acceptable_fits} we show a corner plot with false color
indicating the number of acceptable best fit simulations having
particular parameter combinations. Based on the plot, we find
acceptable fits for the parameter ranges $-1.6 \le a \le -1.2$, $-3.2
\le \gamma \le -2.8$, $7.5 \le k \le 8$, and $-0.35 \le \eta \le
-0.30$.

We constructed two additional simulation grids to investigate whether
better fits could be obtained using a fixed value of $\alpha = +0.22$
(corresponding to a Doppler boosting index of $p = 1.78$), and $\alpha
= -0.5$ (corresponding to $p = 2.5$).  The best fit simulation in the
$p = 1.78$ case (model B in Table~\ref{bestfitMC}) gave acceptable
fits to the flux density, redshift, and luminosity distributions, but
provided a relatively poor fit to the apparent speed distribution.
Although the best fit simulation in the $p=2.5$ grid (model C)
provided a good fit to the apparent speed distribution, none of the
simulations in the grid gave A-D probabilities greater than 0.03 in
all four observable parameters simultaneously.

\begin{comment}
% From best fit model p2large/largebestmod.dat, 1000 subtrials
gidx=-1.40_gl=1.25000_gu=50_xi=0.0_L1=24.00_LFidx=-3.10_zl=0.150_zu=3.4_k=8,00_eta=-0.35

Var        mean   sigma 
S          0.62   0.26 
z          0.58   0.29 
P          0.45   0.25 
Bapp       0.32   0.19 
P - Bapp   0.37   0.22 
P-b_chsq  89.39   8.93 
\end{comment}

\begin{deluxetable*}{lllllrrcccc}
\tablecolumns{11} 
\tablewidth{0pt}  
\tablecaption{\label{bestfitMC} Best Fit Monte Carlo Grid Simulations}  
\tablehead{\colhead{Model} & \multicolumn{6}{c}{Parameter}& \multicolumn{4}{c}{Anderson-Darling Test Probabilities}  \\
\colhead{Name} & \colhead{$b$} & \colhead{$\gamma$} & \colhead{$k$} &
\colhead{$\eta$} &  \colhead{$\alpha$} & \colhead{$p$} & \colhead{$p_S$} & \colhead{$p_z$} & \colhead{$p_P$} & \colhead{$p_{\beta\mathrm{app}}$}  }
\startdata 
A & $-$1.40 & $-$3.1 &8.0& $-$0.35& 0    & 2   & 0.65  & 0.43 & 0.52 & 0.40 \\
B & $-$1.00 & $-$2.6 &8.5& $-$0.30&0.22  & 1.78& 0.70  & 0.67 & 0.24 & 0.10 \\
C & $-$1.40 & $-$3.2 &7.0& $-$0.35&$-$0.5& 2.5 & 0.13  & 0.044 & 0.026 & 0.46
\enddata 
\tablecomments{The simulation parameters are defined in Table 6. Model A has the highest overall Anderson-Darling test probability sum   $p_S$+ $p_z$+ $p_P$+ $p_{\beta\mathrm{app}}$ of any grid simulation.}
\end{deluxetable*}

\subsubsection{\label{bestfitsim}Best Fit Parent Population Properties}

\begin{figure*}
\begin{center}
\includegraphics[angle=0,trim=1cm 0cm 1cm 1.5cm,clip,width=0.9\textwidth]{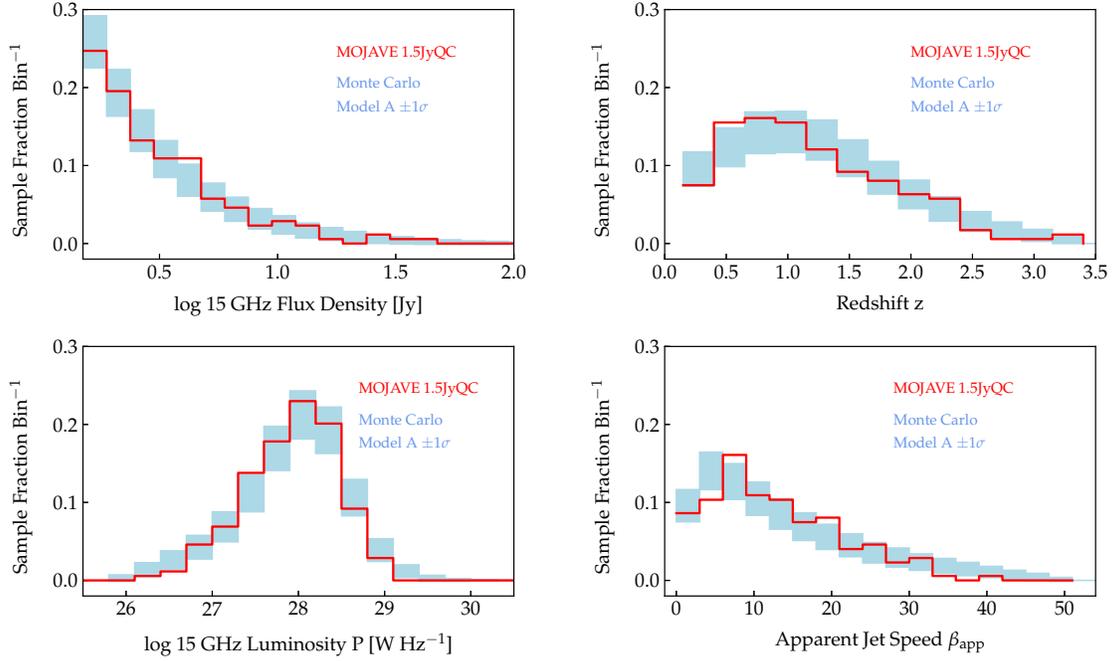}
\end{center}
\caption{\label{sim_hist1} Histograms of observable jet
  properties for the MOJAVE 1.5JyQC quasar sample (red lines). The
  blue bands indicate the $\pm 1\sigma$ ranges on the bin values
  obtained by drawing 10000 samples of 174 jets from the best fit Monte
  Carlo simulation.}
% Produced by 
% running fluxsamp_print_all_gammas with nsamp = 100*174 to create gammas.out
% run histogram_sigmas.pl
% then run ngammaplot.py, take output and modify histplot.py (update
% parentgammas array)
% using histogram_sigmas.pl  and histplot.py
\end{figure*}

In Figure~\ref{sim_hist1} we show the distributions of observable
quantities for the 1.5JyQC quasar sample (red lines), as well as our
best fit (model A) simulation. The blue bands represent 1$\sigma$
ranges on the bin values that we derived by producing a simulation
1000 times the size of the 1.5JyQC, and then randomly choosing a
sub-sample of 174 jets from it, repeating the latter step 10000 times.
We note that the simulation plotted in Fig.~\ref{sim_hist1} provides
the best overall fit to the data, however, other combinations of fit
parameters gave better fits to individual observable quantities. We
have scaled the simulated apparent speed distribution in the top right
panel by a factor of $151/174 = 0.87$ to take into account the 23
missing jet speeds in the 1.5JyQC quasar sample.

\begin{figure*}
\begin{center}
\includegraphics[angle=0,trim=1.5cm 0cm 2cm 1.5cm,clip,width=0.95\textwidth]{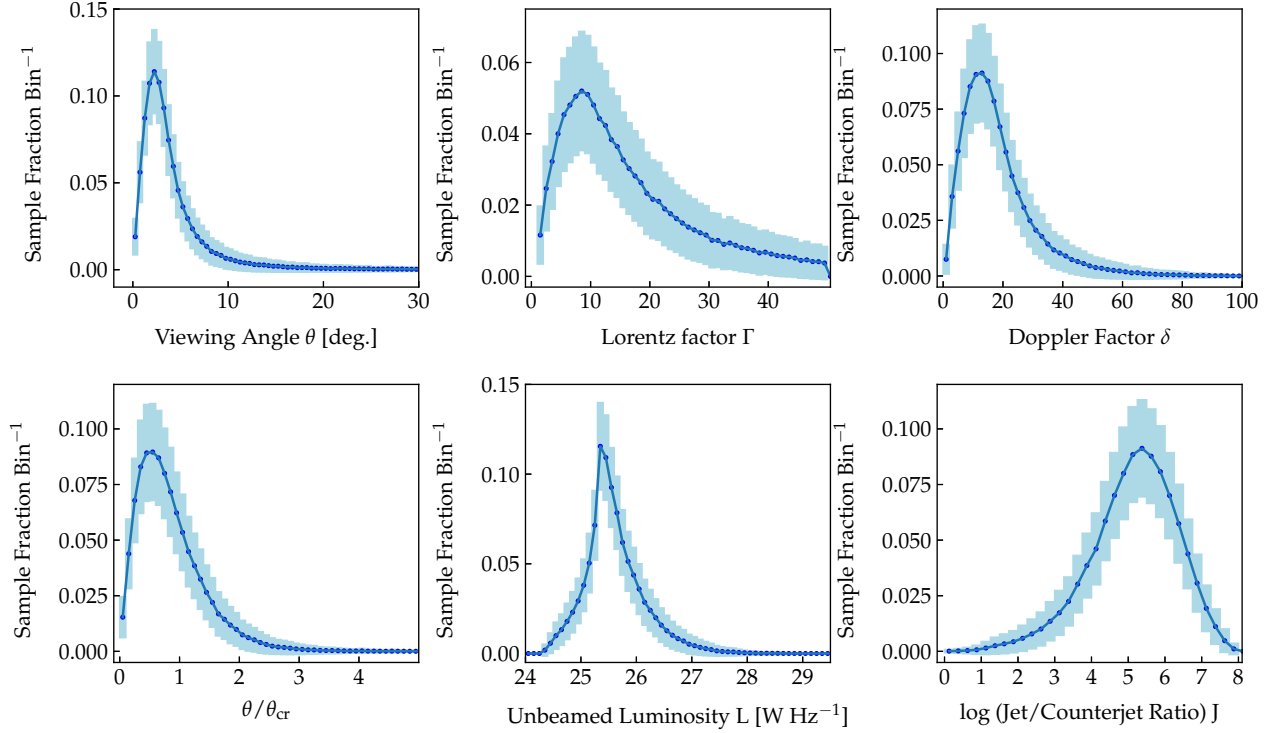}
\end{center}
\caption{\label{sim_hist2} Histograms of intrinsic jet
  properties for the best fit Monte Carlo simulation of the 1.5JyQC
  quasar sample. The blue bands indicate the $\pm 1\sigma$ ranges on
  the bin values obtained by drawing 1000 samples of 174 jets from the
  simulation.}
\end{figure*}

We plot the distributions of several intrinsic (indirectly observable)
quantities from our best fit simulation A in Figure~\ref{sim_hist2}. As
expected from Doppler orientation bias, nearly all of the quasar jets
in the 1.5JyQC sample are predicted to have viewing angles less than
$\sim 10^\circ$ from the line of sight, with the distribution peaking
at $2^\circ$.  The bottom left panel shows the distribution in terms
of the critical angle $\theta_\mathrm{cr} = \sin^{-1}{(1/\Gamma)}$,
and indicates that the most likely viewing angle is not
$\theta_\mathrm{crit}$ as commonly cited in the literature, but
approximately half of this value (e.g., \citealt{VC94, LM97,
  2007ApJ...658..232C}). The top middle panel shows the Lorentz factor
distribution, which is broadly peaked between $\Gamma \simeq 5 $ and
$\Gamma \simeq 15$, with a rapid falloff past $\Gamma = 20$.  The
breadth of the $\Gamma$ distribution indicates that adopting a single
value of $\Gamma = 10$ for all blazars is not well supported by the
observational data. The Doppler factor distribution has a similar
shape to the $\Gamma$ distribution, and peaks at $\delta \simeq 10$,
declining rapidly past $\delta \simeq 30$.

\cite{2018ApJ...866..137L} recently carried out a Bayesian light curve
analysis of OVRO 15 GHz monitoring data on the original 1.5 Jy sample
and calculated variability Doppler factors and distributions of
Lorentz factor and viewing angle. We find a high degree of consistency
between these distributions for the 1.5 Jy quasars and those of our
best fit Monte Carlo simulation in Fig.~\ref{sim_hist2}.

In the bottom middle panel we plot the distribution of intrinsic
(unbeamed) luminosity $L$. Although the intrinsic parent LF peaks at
$L\simeq 10^{24}$ W Hz$^{-1}$, most of the jets in the simulated flux
density-limited sample have intrinsic (unbeamed) luminosities roughly
an order of magnitude higher due to the combined effects of Doppler
and Malmquist bias. This implies that the parent population of the
brightest radio quasars consists of powerful FR II radio galaxies with
a relatively narrow range of unbeamed 15 GHz radio luminosity between
$\sim 10^{25}$ and $\sim 10^{26}$ W Hz$^{-1}$.

\cite{2017MNRAS.465..180L} used Monte Carlo simulations to investigate
the predicted distribution of jet-counterjet flux density ratios due
to relativistic beaming in flux density-limited blazar samples. In the bottom
right panel of Fig.~\ref{sim_hist2} we plot the distribution of this
quantity for our best fit model. We obtain very similar results, with
most jets having ratios of $10^4$ to $10^7$. These are much higher
than can be probed in our snapshot MOJAVE VLBA images, given their
image rms levels of $\sim 0.1$ mJy beam$^{-1}$ and typical jet
brightnesses of $<100$ mJy beam$^{-1}$ downstream from  the core.

\begin{figure}
\begin{center}
\includegraphics[angle=0,trim=0cm 0cm 1cm 0cm,width=0.95\columnwidth]{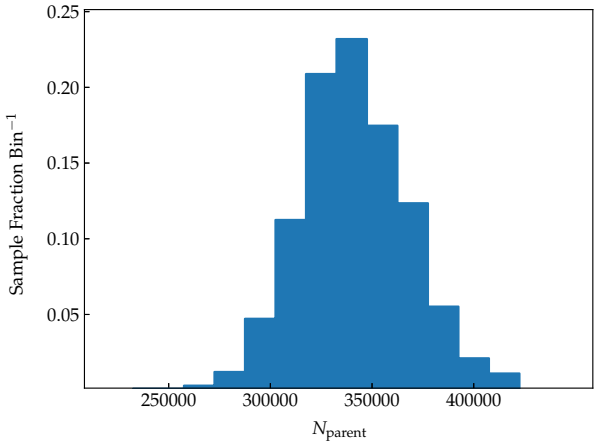}
\end{center}
\caption{\label{sim_npop} Distribution of parent population
  size in the best fit Monte Carlo simulation that is required to
  produce each of 1000 sub-samples of 174 jets matching the MOJAVE
  1.5JyQC properties.}
\end{figure}

In Figure~\ref{sim_npop} we plot the distribution of parent population
sizes for the 10000 sub-samples, which is approximately Gaussian. For
our best fit simulation parameters, typically $(3.5 \pm 0.3) \times
10^5$ parent jets are needed to reproduce the 174 quasar jets in the
MOJAVE 1.5JyQC sample. Given the co-moving simulated volume of 1334
Gpc$^3$, this implies a parent space density of $261 \pm 19$
Gpc$^{-3}$, which is comparable to the value of 200 Gpc$^{-3}$
obtained for FR II radio galaxies by \cite{2001MNRAS.328..897S} using
the LF of \cite{1990MNRAS.247...19D}.

A rule of thumb sometimes used in the literature is that for every
blazar jet found in a survey with Lorentz factor $\Gamma$ there are
$\Gamma^2$ parent jets (e.g.,
\citealt{Mutel1990,Ghisellini2000,2016AA...591A..98B}). This
assumption is based on the ratio of solid angle subtended by blazar
jets viewed within the critical angle $1/\Gamma$ and the full range
of jet viewing angle in the parent population, but fails to properly
take into account the biases of flux density-limited sampling.

\begin{figure}
\begin{center}
\includegraphics[angle=0,trim=0cm 0cm 1cm 0cm,width=0.95\columnwidth]{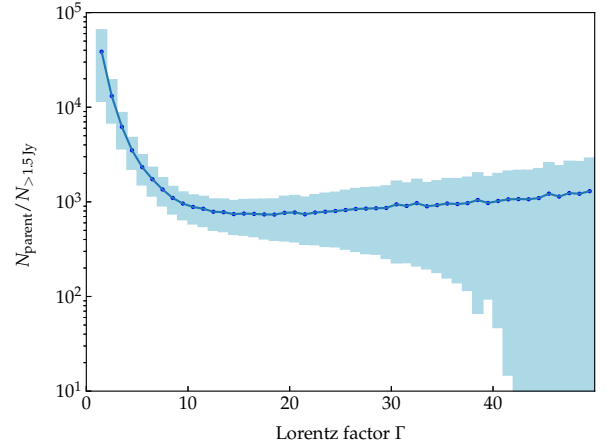}
\end{center}
\caption{\label{gamma_npop}Mean number of jets in the parent
  population divided by the mean number of jets exceeding 1.5 Jy in binned
  intervals of Lorentz factor for the best fit Monte Carlo simulation.
  The blue bands indicate the $\pm 1\sigma$ ranges on the bin values
  obtained by drawing 1000 samples of 174 jets from the simulation.}
\end{figure}

In Figure~\ref{gamma_npop} we plot for our best fit Monte Carlo
simulation the number of parent jets divided by the number of
simulated $S_\nu > 1.5$ Jy jets in binned intervals of Lorentz factor between
1 and 50. For $\Gamma \gtrsim 15$, there is a shallow increase in the
predicted number of parent jets for each jet found with a particular
Lorentz factor in the 1.5JyQC sample, from $N \sim 750$ to $N \sim
1300$ at $\Gamma = 50$. This is much shallower than the rule of thumb
$\Gamma^2$ dependence, and is a result of the fact that (i) very high
$\Gamma$ jets are rare in the parent population, and (ii) most of
these high $\Gamma$ jets do not exceed the 1.5 Jy flux density cutoff
not only due to their viewing angle, but also their redshift and/or
unbeamed luminosity. The large range of possible parent sizes for
$\Gamma > 40$ reflects the statistical fluctuations associated with
selecting from this small cohort of jets in a flux density-limited
sample.

A different behavior is seen below $\Gamma \simeq 15$.  These jets are
abundant in the parent population, yet most have low unbeamed
luminosities and require either substantial Doppler boosting or a low
redshift to exceed the flux density cutoff. Every low $\Gamma$ jet in
the 1.5JyQC requires significantly more parent objects, since its
maximum possible Doppler boost is only $(2\Gamma)^p$.  This is the exact
opposite of the  $N \propto \Gamma^2$ prediction. 

\section{SUMMARY AND CONCLUSIONS}

We have carried out a study of the parsec-scale jet kinematics of 409
bright radio-loud AGNs above declination $-30^\circ$, based on 15 GHz
VLBA data obtained between 1994 August 31 and 2016 December 26. These
AGNs have been part of the 2cm VLBA survey or MOJAVE programs, and
have $\gtrsim 0.1$ Jy of correlated flux density at 15 GHz.  By
modeling the jet emission with a series of Gaussians in the
interferometric visibility plane, we identified and tracked 1744
individual features in 382 jets over at least five epochs. We fitted
their sky trajectories with simple radial and vector motion models,
and additionally carried out a constant acceleration fit for 881
features that had ten or more epochs.

A primary goal of the MOJAVE program is to characterize the jet
properties of a well-defined flux density-limited sample in order to
better understand the blazar parent population. Using the extensive
OVRO and UMRAO single-dish monitoring databases, as well as the MOJAVE
VLBA archive, we constructed the MOJAVE 1.5 Jy Quarter Century sample,
which consists of all 232 AGNs north of J2000 declination $-30^\circ$
that are known to have exceeded 1.5 Jy in 15 GHz VLBA correlated flux
density between 1994.0 and 2019.0. We carried out Monte Carlo
simulations to determine the best fit parent population parameters
that reproduced the redshift, radio luminosity, and apparent velocity
distributions of the 174 quasars with $z \ge 0.15$ in the 1.5JyQC
sample.

We summarize our conclusions as follows:

1. A total of 382 of 409 jets had at least one robust bright feature that
could be tracked for five or more epochs.  A majority (59\%) of the
well-sampled jet features showed evidence of
accelerated motion at the $>3 \sigma$ level. 

2. We examined the distribution of apparent speeds within 26
individual jets that had ten or more robust features, and confirmed that
each jet tends to have a characteristic speed that is likely related
to the underlying flow. Other than a few fast outliers and some slow
pattern speeds, the speeds of features in a jet typically lie within
$\sim \pm 40\%$ of the characteristic speed. 

3. We were able to identify 55 features in 42 jets that had unusually
slow pattern speeds ($\mu < 20$ \muasyr and at least 10 times slower
than the fastest feature in the jet).  We confirm the 43 GHz VLBA
results of \cite{2017ApJ...846...98J} that the vast majority of these
lie within 4 pc (projected) of the core feature, and may represent
quasi-stationary standing shocks near the jet base.

4. Only 2.5\% of the features we studied had velocity vectors directed
inward toward the core. In some cases, these are likely due to
brightness variations affecting the fitted centroid position of a
large diffuse feature, or a feature on a bent trajectory that is
crossing the line of sight. In other cases there may be a
mis-identification of the true core position. We find that in 16 of
the 32 jets with apparent inward motion, the inward-moving feature is
the closest feature to the core, and that four BL Lac jets have more
than one close-in inward-moving feature.

5. We examined the distribution of maximum apparent jet speed for
the AGNs in our full sample and the 1.5JyQC sample, and find that
it is peaked at low values, with very few speeds above 30 $c$.  Given the
fact that large Doppler-biased jet samples should contain examples of
the fastest jets in the parent population, and that our survey has not
measured any instantaneous speeds above 50 $c$, this implies that the
parent distribution of jet Lorentz factors is not single-valued, but is
weighted towards low $\Gamma$, with decreasing numbers of jets up to
$\Gamma_\mathrm{max} = 50$.

6.  We find a strong correlation between apparent jet speed and
synchrotron peak frequency, with the highest jet speeds being found
only in AGNs with low $\nu_p$ values.  Although a fast jet speed does
not guarantee that a jet will be detected at TeV gamma-ray energies,
it appears to be a minimum requirement for LSP and ISP AGNs. The
exceptions to date are the two very nearby radio galaxies 3C 84 and
M87, and the BL Lac TXS 0506+056 that has been associated with a high
energy neutrino detection event.

7. Our large grid of Monte Carlo parent population simulations yielded
several parameter combinations that could adequately reproduce the
flux density, redshift, radio luminosity, and apparent velocity
distributions of the 174 quasars in the 1.5JyQC sample.  These
simulations have an unbeamed luminosity function above $10^{24}$ Hz
with power law slope $-3.2 \le \gamma \le -2.8$, and pure luminosity
evolution of the form $ e(z) = (1+z)^{k} e^{z/\eta}$, where $7.5 \le k
\le 8$ and $-0.35 \le \eta \le -0.30$.  The parent jet population has
a power law distribution of Lorentz factors with slope $-1.6 \le b \le
-1.2$, ranging from $\Gamma = 1.25$ to $\Gamma = 50$, and a Doppler
boosting index $p = 2$. The best fit parent population (with $b =
-1.4$, $\gamma = -3.1$, $k = 8.0$, and $\eta = -0.35$) has a space
density of $261 \pm 19$ Gpc$^{-3}$, which is consistent with that of
FR II radio galaxies.  Most of the quasars in the 1.5JyQC have a
relatively narrow range of intrinsic (unbeamed) parsec-scale 15 GHz
radio luminosity between $\sim 10^{24.5}$ W Hz$^{-1}$ and $\sim
10^{26.5}$ W Hz$^{-1}$.

8.  Our best fit simulation indicates that nearly all of the 1.5JyQC
quasar jets are viewed at less than $\sim 10^\circ$ from the line of
sight, with the distribution peaking at $2^\circ$.  As previously
discussed by \cite{VC94, LM97} and \cite{2007ApJ...658..232C}, the
most probable jet viewing angle is $\sim$ 0.5 times the critical angle
$\theta_\mathrm{cr}= \sin^{-1}(1/\Gamma)$ where $\beta_\mathrm{app} = \Gamma\beta$.

9. The Lorentz factor distribution of the 174 bright radio quasars in
the flux density-limited 1.5JyQC sample peaks between $\Gamma = 5$ and
$\Gamma = 15$, with a rapid falloff past $\Gamma = 20$.  The breadth
of the $\Gamma$ distribution indicates that adopting a single value of
$\Gamma = 10$ for all blazars is not well supported by the
observational data.  The Doppler factor distribution has a similar
shape to the $\Gamma$ distribution, and peaks at $\delta \simeq 10$,
declining rapidly past $\delta \simeq 30$. Both distributions are
similar to those inferred from variability Doppler factor estimates
using OVRO 15 GHz monitoring data by \cite{2018ApJ...866..137L}.

10. We find that the oft-cited rule of thumb that for every jet found
in a survey with Lorentz factor $\Gamma$ there are $\Gamma^2$ parent
jets is incorrect for flux density-limited blazar samples. Above $\Gamma
\simeq 15$, there is only a shallow increase in the expected number of
parent jets per source with $\Gamma$, while for lower Lorentz
factors, the number of parent jets increases rapidly with decreasing
$\Gamma$.

%%%%%%%%%%%%%%%%%%%%%%%%%%%%%%%%%%%%%%%%%%%%%%%%%%%%%%%%%%%%%%%%%%%%%%

\acknowledgments

The MOJAVE project was supported by NASA-{\it Fermi} grants
NNX08AV67G, NNX12A087G, and NNX15AU76G.  YYK and ABP were supported by
the Basic Research Program P-28 of the Presidium of the Russian
Academy of Sciences and the government of the Russian Federation
(agreement 05.Y09.21.0018).  TS was supported by the Academy of
Finland projects 274477, 284495, and 312496. TH was supported by the
Academy of Finland project 317383.  The Very Long Baseline Array and
the National Radio Astronomy Observatory are facilities of the
National Science Foundation operated under cooperative agreement by
Associated Universities, Inc. This research has used observations with
RATAN-600 of the Special Astrophysical Observatory of the Russian
Academy of Sciences.  This work made use of the Swinburne University
of Technology software correlator \citep{2011PASP..123..275D},
developed as part of the Australian Major National Research Facilities
Programme and operated under licence.  This research has made use of
data from the OVRO 40-m monitoring program \cite{2011ApJS..194...29R},
which is supported in part by NASA grants NNX08AW31G, NNX11A043G, and
NNX14AQ89G and NSF grants AST-0808050 and AST-1109911. This research
has made use of data from the University of Michigan Radio Astronomy
Observatory which has been supported by the University of Michigan and
by a series of grants from the National Science Foundation, most
recently AST-0607523.  This research has made use of the NASA/IPAC
Extragalactic Database (NED) which is operated by the Jet Propulsion
Laboratory, California Institute of Technology, under contract with
the National Aeronautics and Space Administration.

\vfill\eject
\appendix{\bf Appendix: Notes on Individual AGNs}

Here we provide comments on individual AGNs supplementing those given
in \cite{MOJAVE_X} and \cite{MOJAVE_XIII}. 

0111+021 (UGC 00773): All five features closest to the core in this
nearby BL Lac jet ($z = 0.047$) have inward or possibly
inward-directed motions.

0118$-$272 (OC $-$230.4): New Gaussian fitting to the epoch data
indicates that feature id = 3 no longer has inward motion at the
$>3\sigma$ level as was reported by \cite{MOJAVE_XIII}.

0256+075 (OD 94.7): The large time gaps between epochs made it
impossible to reliably cross-identify any robust features in this
quasar.

0300+470 (4C +47.08) : Additional new epochs and a re-analysis of the
data indicates that the previously reported inward-moving feature (id
= 2) in this AGN is not robust.  The redshift for this BL Lac is
unknown, with the NED value of $z = 0.475$ being an arbitrary
assignment.  \cite{2013ApJ...764..135S} find $0.37 < z < 1.63$ based
on an optical spectrum.

0518+211 (RGB J0521+212): The two innermost jet features in this BL
Lac object have statistically significant inward motion.

0710+196 (WB92 0711+1940): The jet features in this quasar were too
weak ($<$ 10 mJy) to identify as robust.

1101+384 (Mrk 421): All three innermost jet features of this nearby BL Lac
object show inward motion. 

1118+073 (MG1 J112039+0704): The location of the core in this jet is
uncertain. We assumed the core to lie at the northeasternmost point in
the jet.

1148$-$001 (4C $-$00.47): We identified the core as the most compact
feature of the jet, with a 2 mas feature (id = 5) being located upstream. 

1215+303 (ON 325): All three innermost features of this low redshift
BL Lac object ($z = 0.131$) show small but significant inward motion
of approximately 25 $\muasyr$ (0.2 c).

1224$-$132 (PMN J1226$-$1328): We were unable to identify any robust
jet features in this BL Lac object.

PG 1246+586: None of the jet features in this BL Lac object were
sufficiently bright or compact enough to be identified as robust.

1253$-$055 (3C 279): The VLBA epochs in 2013--2014 are affected by the
emergence of two very bright features ($> 10$ Jy).  The most
consistent fits during 2014--2015 required fitting an upstream feature
(id = 17) that could be the true core that is only strong enough to be
visible during these epochs.  The reference 'core' position that we
use in all of our fits may thus be a strong quasi-stationary feature in the
flow.

1300+248 (VIPS 0623): There was no jet feature in this BL Lac object
that was sufficiently bright or compact enough to be identified as
robust.

PKS 1402+044: The innermost jet feature (id = 5) of this quasar shows
statistically significant inward motion.

1458+718 (3C 309.1): A re-analysis of the complex located 23 mas south
of the core now indicates no significant inward motion.

PG 1553+113: We were unable to identify any robust jet features in
this BL Lac object.

1557+565 (VIPS 0926): The NED redshift of z = 0.3 is not confirmed by
\cite{2013AJ....146..127S}, who find a lower limit of $z > 1.049$. The
innermost jet feature (id = 4) of this BL Lac object shows
statistically significant inward motion.
 
1656+482 (4C +48.41): The innermost jet feature (id = 4) of this BL
Lac object shows statistically significant inward motion.

TXS 1811+062: We were unable to identify any robust jet features in
this BL Lac object.

1928+738 (4C $+$73.18): In 2012 a counter-jet feature with an apparent
speed of 0.8 $c$ emerged in this quasar jet.

8C 1944+838: The outermost feature (id = 1) in this jet has a
statistically significant inward speed, but may not represent true
motion due to the large, diffuse nature of the emission. 

1ES 1959+650: The jet structure was too weak and compact at 15 GHz to
reliably measure any robust features. \cite{2010ApJ...723.1150P}
obtained a maximum speed measurement of $0.0322 \pm 0.0064$ \masyr at
43 GHz.

2028+492 (MG4 J202932+4925): We were unable to identify any robust jet
features in this BL Lac object.

2234+282 (CTD 135): An unpublished 43 GHz VLBA image by Tao An
suggests that core is located in the southwest portion of the jet. 

TXS 2308+341: The brightest feature in this jet does not appear to be
a stable reference point, with correlated positional changes seen in
the positions of downstream features seen at several epochs.

S5 2353+816: No robust jet features could be identified in this BL Lac
object due to the large time gap in the dataset.

\bigskip
\bibliographystyle{apj}
\bibliography{lister}

\end{document}